\documentclass[article,aps,a4paper,twocolumn]{revtex4-2}
\usepackage{fullpage} 
\usepackage{parskip} 
\usepackage{tikz} 
\usepackage{amsmath,amsfonts}
\usepackage{array}
\usepackage{mathtools}
\usepackage{hyperref}
\usepackage{dsfont}
\usepackage[normalem]{ulem}
\usepackage{newunicodechar}

\newenvironment{myquote}%
  {\list{}{\leftmargin=0.3in\rightmargin=0.3in}\item[]}%
  {\endlist}

\newcommand{\ad}[1]{\textrm{ad}_{#1}}
\newcommand{\E }{E_f}
\newcommand{\tr}{\textrm{tr}}
\newcommand{\dt}{\ \cdot \ }

\begin{document}

\title{The classical-quantum limit}

\author{Isaac Layton}
\author{Jonathan Oppenheim}
\affiliation{Department of Physics and Astronomy, University College London, Gower Street, London WC1E 6BT, United Kingdom}

\begin{abstract}
The standard notion of a classical limit, represented schematically by $\hbar\rightarrow 0$, provides a method for approximating a quantum system by a classical one. In this work we explain why the standard classical limit fails when applied to subsystems, and show how one may resolve this by explicitly modelling the decoherence of a subsystem by its environment. Denoting the decoherence time $\tau$, we demonstrate that a double scaling limit in which $\hbar \rightarrow 0$ and $\tau \rightarrow 0$ such that the ratio $E_f =\hbar /\tau$ remains fixed leads to an irreversible open-system evolution with well-defined classical and quantum subsystems. The main technical result is showing that, for arbitrary Hamiltonians, the generators of partial versions of the Wigner, Husimi and Glauber-Sudarshan quasiprobability distributions may all be mapped in the above double scaling limit to the same completely-positive classical-quantum generator. This provides a regime in which one can study effective and consistent classical-quantum dynamics.

\end{abstract}

\maketitle

\section{Introduction}

The classical limit describes the emergence of classical physics from quantum theory. Typically justified in a variety of ways, the most famous of these is to consider action large compared to the reduced Planck's constant, $\hbar$. This leads to the ubiquitous statement that the classical limit is taking $\hbar \rightarrow 0$. As well as explaining the success of classical mechanics for the description of macroscopic systems, the classical limit provides an important theoretical tool for simplifying the analysis of quantum systems that are too complex to study directly.

The classical limit allows one to replace a quantum system with an entirely classical description. However, many systems 
of interest operate in a regime where both classical and quantum features are important, and this leads to the following question:

\begin{myquote}{}
    Can we take a limit of a quantum system such that one subsystem behaves classically, while the rest remains quantum?
\end{myquote}

A limit of this kind would have a wide variety of applications: from providing first principle derivations of quantum control and measurement set-ups \cite{ezawa1993quantum,lloyd2000coherent,wiseman_milburn_2009}; to describing systems at the classical-quantum boundary \cite{bertet2001complementarity,PhysRevLett.89.018301,PhysRevLett.130.193603}; formalising approaches in quantum chemistry where the nuclear degrees of freedom are treated as classical and the electronic degrees of freedom are treated quantum mechanically \cite{tully1990molecular,tully1998mixed,kapral1999mixed}; and more generally providing a framework to simplify complex many-body quantum dynamics while retaining essential quantum features. Beyond these more practical applications, it would be interesting if recently proposed models of classical-quantum theories of gravity \cite{oppenheim2023postquantum,oppenheim2023gravitationally,layton2023weak,Galley2023anyconsistent,oppenheim2023covariant,oppenheim2024diffeomorphism,weller2024quantum} could arise as effective descriptions of quantum gravity.

In this work, we tackle the problem of taking such a classical limit. Since this involves mapping two quantum subsystems to a quantum subsystem and an effective classical subsystem, we call this a ``quantum-quantum to classical-quantum" limit, or classical-quantum limit for short. Such a limit could also be referred to as a semi-classical limit, since the resulting effective theory contains both classical and quantum degrees of freedom, or as a classical limit for subsystems.  We use this terminology in a more general sense than earlier work – this limit is taken to be able to include the effects of \textit{back-reaction} on the effective classical system, and therefore should not be conflated with special cases in which the effective classical system is unaffected by the quantum one \cite{oliynyk2016classical,irish2022defining}. 

Two important requirements to make on any classical-quantum limit of quantum theory is that it be physically motivated and consistent. Although the standard $\hbar \rightarrow 0$ classical limit is often well-motivated physically, we shall see that as a classical-quantum limit it fails to be consistent, in that it fails to describe an effective classical subsystem. In this case, the resulting dynamics is known as the quantum-classical Liouville equation \cite{Aleksandrov+1981+902+908,gerasimenko1982dynamical,kapral1999mixed}, and does not lead to well-defined classical evolution on the subsystem in question \cite{boucher1988semiclassical}. The first example of a limit procedure leading to consistent dynamics was provided in the pioneering work by Di\'osi \cite{diosi1995quantum} who considered two particles, each having a different Planck's constant. 
While this allowed for one of the first examples of consistent classical-quantum dynamics to be derived, it came at the expense of requiring a number of unphysical considerations, including modified quantum mechanical evolution laws and ad hoc sources of classical noise. 

In the present work we demonstrate that a physically motivated and consistent limit procedure exists, starting from standard unitary quantum mechanics in a closed system. The key observation we make is the following: such closed system dynamics will always generate entanglement between subsystems, and thus always lead to a breakdown of classicality, independently of any parameter such as $\hbar$ that is usually used to quantify classicality. This means that the standard notions of a classical limit, such as $\hbar \rightarrow 0$, must be supplemented by an additional mechanism that removes the entanglement generated between subsystems.

In our framework, the classicality of a subsystem is guaranteed by decoherence due to its external environment. Already well understood to play an important role in the quantum-to-classical transition \cite{zurek1994decoherence,schlosshauer2005decoherence}, the coupling to the environment in our framework leads to an associated decoherence timescale $\tau$ of the subsystem in question. By taking $\tau\rightarrow 0$, one can ensure that this subsystem is classical at all times. The key conceptual takeaway from this work is that a double scaling limit, in which $\hbar \rightarrow 0$ while $\tau\rightarrow 0$ such that the ratio $\E=\hbar/\tau$ is fixed, provides a version of a classical limit that may be consistently applied to subsystems i.e. a classical-quantum limit. 

The main technical result of this work is computing the explicit form of the dynamics in this double scaling limit, for arbitrary bipartite Hamiltonians for which a classical limit is possible, which is given in Equation \ref{eq: LcanCQ}. In order to prove the consistency of this dynamics, we show that the dynamics is a special case of the recently characterised completely-positive form \cite{oppenheim2023postquantum,oppenheim2022two}, which guarantees that the effective classical subsystem is well-defined. This distinguishes our dynamics from earlier attempts at deriving effective classical-quantum dynamics from quantum theory \cite{HalliwelDH,diosi2000quantum}, and provides a regime in which the general form of continuous dynamics introduced in  \cite{oppenheim2022two} could arise as an effective theory. The resulting dynamics is generically an irreversible open-system dynamics, with decoherence and diffusion controlled by the parameter $\E$. The complete-positivity ensures the classical-quantum dynamics may be directly unravelled in terms of continuous classical trajectories in phase space and quantum trajectories in Hilbert space \cite{layton2022healthier}, which are given in Equations \eqref{eq: unrav_C} and \eqref{eq: unrav_Q}.

Alongside the main results, we find a number of related results:
\begin{itemize}
    \item A partial version of the Glauber-Sudarshan quasiprobability distribution is introduced, and identified as the correct representation to require positivity of for effective classical-quantum dynamics in the Hilbert space.
    \item The dynamics of partial versions of the Glauber-Sudarshan and Husimi quasiprobability distribution are explicitly computed to $O(\hbar^0)$.
    \item The classical-quantum limit is shown to lead to dynamics independent on the choice of partial $Q$, $P$ or $W$ representation.
    \item The double scaling limit applied to a single system is shown to give a stochastic classical limit, of the kind described by \cite{werner1995classical}.
    \item A simplified form of dynamics is found in Equation \eqref{eq: Lapprox} for classical-quantum Hamiltonians that self-commute, which takes the form of $O(\hbar^0)$ partial Glauber-Sudarshan dynamics with the minimal additional decoherence and diffusion for complete-positivity.
    \item The explicit form of dynamics for the classical-quantum limit of two quantum harmonic oscillators is computed.
    \item The double scaling limit on a single system is shown to recover the standard $\hbar \rightarrow 0$ limit in the low diffusion limit $\E\rightarrow 0$. 
    \item Two distinct behaviours of the effective classical-quantum dynamics are found in the $\E\rightarrow 0$ limit, namely a quantum Zeno type behaviour, and a coherent quantum control limit.
\end{itemize}

The results in this paper establish that the classical limit of a subsystem has a far richer structure than the classical limit of a single system, suggesting a new possible arena in which to apply methods developed to study the quantum-to-classical transition \cite{zurek1994decoherence,bhattacharya2000continuous,hernandez2023decoherence}.  We also provide a number of technical tools and definitions for the study of effective classical-quantum theories, which may be useful for categorising the large body of existing proposals for constructing hybrid theories \cite{sudarshan1976interaction,peres2001hybrid,HalliwelDH,diosi2000quantum,hall2005interacting,barcelo2012hybrid,bondar2019koopman, PhysRevD.98.065002,barchielli2023hybrid,gonzalez2023mixed}. Our findings also make clear that effectively classical systems are indeed consistent with quantum theory, and in fact may be derived from it, contrary to early arguments \cite{cecile2011role,dewitt1962definition,eppley1977necessity}.

Aside from being of foundational interest, the form of dynamics we find in equation \eqref{eq: LcanCQ} or their approximated form \eqref{eq: Lapprox} may be directly applied to study a wide range of quantum Hamiltonians in the classical-quantum limit. By using the stochastic unravellings of equations \eqref{eq: unrav_C} and \eqref{eq: unrav_Q}, one may numerically study the statistics of various observables of quantum systems in a regime where subsystems are effectively classical.  This provides a general framework for studying the effects of backreaction from quantum systems on effectively classical ones, as well as a tool for reducing the complexity of simulations of many-body quantum dynamics in the presence of strongly decohering environments, which we anticipate could have a wide variety of applications.

The structure of the paper is as follows: in Section \ref{sec: wig} we introduce the Wigner and partial Wigner representations, and demonstrate using the latter how the $\hbar \rightarrow 0$ limit is insufficient in providing a classical limit of a subsystem. In Section \ref{sec: env} we introduce a discrete-time, decoherence channel model of an environment, and show how this leads to well-defined stochastic evolution in a double scaling limit. In Section \ref{sec: main} we present our main result, which is the derivation of the general form of classical-quantum dynamics under a bipartite Hamiltonian, under this double scaling limit. In Section \ref{sec: technic} we introduce two other partial quasiprobability representations, the partial Glauber-Sudarshan and partial Husimi representations. These are used to illustrate two technical notions useful for characterising effective classical-quantum dynamics, which we use to determine that the positivity of the partial Glauber-Sudarshan distribution is a sufficient and necessary measure of the effective classicality of a subsystem. In Section \ref{sec: quasi} we study the main form of dynamics in the three different quasiprobability distributions introduced, and show the equivalence between them. In Section \ref{sec: traj} the main results are unravelled in terms of stochastic trajectories in phase space and Hilbert space. Finally, in Sections \ref{sec: two} and \ref{sec: E}, special cases of the general form of dynamics are given, in particular the self-commuting classical-quantum Hamiltonian case, and the low diffusion $\E\rightarrow 0$ limit.

\section{ The standard $\hbar \rightarrow 0$ limit } \label{sec: wig}
To motivate the need for an alternate notion of a classical limit, we first begin by looking at where the standard classical limit succeeds and fails as a technique for deriving classical equations of motion. To do so, we will look at the simplest model of a quantum system with a classical limit, i.e. a single quantum system characterised by a canonical commutation relation with parameter $\hbar$. However, the results that follow may be reinterpreted in the standard way for general order parameters controlling the degree of classicality, such as coupling strength $g$ or number of systems $N$ \cite{yaffe1982large}.

Consider a single quantum system, that we denote $C$, with Hilbert space $\mathcal{H}^C$ and trace-one, positive semi-definite density operators $\hat{\rho}$ that form a set of states $S(\mathcal{H}^C)$. We will take this quantum system to be characterised by the canonical commutation relation $[\hat{q},\hat{p}]=i\hbar$, and interpret the operators $\hat{q}$ and $\hat{p}$ as the position and momentum of the system. This system will have an associated Hamiltonian $\hat{H}$ which generates the free, unitary evolution of the $C$ system in the absence of any interactions with other systems.

A typical method of studying the classical limit of such a system is via the Wigner representation, which provides an alternate description of quantum mechanics in terms of functions of phase space \cite{wigner1932quantum,wigner1932quantum,groenewold1946principles,moyal1949quantum,zachos2005quantum}. Defining the operators $\hat{A}_{q,p}=\frac{1}{\pi\hbar}\int dy \ e^{-ip y/\hbar} |q-\frac{1}{2} y\rangle \langle q + \frac{1}{2} y |$, where $|q\rangle$ denotes the eigenstates of the position operator $\hat{q}$, one may map operators acting on $\mathcal{H}^C$ to functions of phase space $\mathcal{M}$ by taking the trace with respect to $\hat{A}_{q,p}$ i.e. $\hat{O}\mapsto\tr[\hat{A}_{q,p}\hat{O}]$. The most important example is the Wigner function $W(q,p)$, the phase space representation of the quantum state 
\begin{equation}
W(q,p)=\tr[\hat{A}_{q,p} \hat{\rho}].
\end{equation}
By the properties of $\hat{A}_{q,p}$ and $\hat{\rho}$, it follows that $W(q,p)$ is real-valued and is normalised when integrated over phase space i.e. $\int W(q,p)\  dq dp=1$. Unlike a probability distribution, it is not guaranteed to be non-negative for all $q,p$ in phase space, and hence is termed a \textit{quasi}probability distribution. To study how unitary dynamics are represented in the Wigner representation, one must also consider the Wigner representation of the Hamiltonian $\hat{H}$, given by the real-valued $H^W(q,p)=\tr[{\hat{A}_{q,p} \hat{H}}]$. The free unitary dynamics in the Wigner representation is then given by the Moyal bracket 
\begin{equation} \label{eq: WigEvo}
    \frac{\partial W}{\partial t}=-\frac{i}{\hbar}(H^W\star W - W\star H^W),
\end{equation} where here the star product of two phase space functions $f=f(q,p)$ and $g=g(q,p)$ is given
\begin{equation} \label{eq: star}
    f \star g = f\  e^{\frac{i \hbar}{2} (\overset{\leftarrow}{\partial}_q \overset{\rightarrow}{\partial}_p - \overset{\leftarrow}{\partial}_p \overset{\rightarrow}{\partial}_q )}\  g,
\end{equation} and to be interpreted in terms of the series expansion of the exponential, with the arrows denoting whether each derivative acts on the function on the left or the right. Expressions such as equation \eqref{eq: WigEvo} may be found by using the standard result that products of operators on the Hilbert space are mapped to the star product of their Wigner representations i.e. $\hat{f} \hat{g} \mapsto f\star g$ \cite{groenewold1946principles,zachos2005quantum}.

The Wigner representation is an entirely equivalent description of quantum mechanics, and does not \textit{a priori} have anything to do with classical dynamics. However, by considering the dynamics to lowest order in $\hbar$, one arrives at an equation familiar from classical mechanics. Specifically, to lowest order in $\hbar$, the dynamics \eqref{eq: WigEvo} takes the form
\begin{equation} \label{eq: C_liouville}
\begin{split}
    \frac{\partial W}{\partial t}=\{H,W\},
\end{split}
\end{equation} where $\{\dt,\dt\}$ denotes the Poisson bracket, and $H$ denotes the classical Hamiltonian i.e. the $O(\hbar^0)$ part of $H^W$ \cite{yaffe1982large}. This equation is the Liouville equation, and describes how classical probability distributions evolve under Hamiltonian flow. This leads to the statement that $\hbar \rightarrow 0$ gives the classical limit of a quantum system. Of course, it is not actually meaningful to send a dimensionful quantity like $\hbar$ to zero, although it is often a convenient shortcut in practice. The statement that classical equations of motion are recovered in the $\hbar\rightarrow 0$ limit is more precisely understood as the statement that for a given $W(q,p)$, the higher order derivatives terms containing $\hbar$ in the expansion are negligible compared to the Liouville equation terms given above. 

Even if one did not already know the form of the Liouville equation, one is still led to the $\hbar\rightarrow0$ limit by considering when the dynamics preserves the \textit{classicality} of initial states. A full definition of effective classicality will be given in section \ref{sec: technic}, but for the time being we will simply state that for a quantum state $\hat{\rho}$ of a single system to be viewed as effectively classical, it is necessary for the corresponding Wigner function to be positive i.e. 
\begin{equation} \label{eq: W+}
    W(q,p)\geq 0 \quad \forall q,p\in \mathcal{M}.
\end{equation} Correspondingly, for the dynamics of the $C$ system to be effectively classical, it must also be positive i.e. preserve the positivity of all normalised functions of phase space. As should be expected, the general quantum dynamics of equation \eqref{eq: WigEvo} is not positive, except in the cases that the Hamiltonian is at most quadratic in $q$ and $p$. To see this, one may appeal to the Pawula theorem, which characterises the general form of linear, trace-preserving and positive dynamics for real-valued functions of phase space \cite{pawula1967generalizations}. The Pawula theorem states that unless the dynamics contains an infinite number of higher order derivatives with respect to $q$ and $p$, any positive dynamics must be of the Fokker-Planck form, with at most second order derivatives in $q$ and $p$ (see Appendix \ref{app: Pawula} for more details). Since the series expansion of \eqref{eq: WigEvo} typically truncates at a finite number of terms (i.e. for Hamiltonians polynomial in position and momentum), the dynamics in such cases cannot be positive. 

Considered in this way, the $\hbar\rightarrow 0$ limit may be understood as a method of enforcing positivity preservation on the quantum dynamics of a single system when represented in phase space. In particular, since the higher order derivative terms in equation \eqref{eq: WigEvo} responsible for violating positivity also are higher order in $\hbar$, by truncating the expansion to lowest order in $\hbar$ the dynamics reduces to a dynamics that maps initial probability distributions to final probability distributions, and hence preserves the classicality of initial states.

However, let us now consider the same approach when the $C$ system is just a \textit{subsystem} of a larger quantum system. Denoting the other subsystem $Q$, we again denote states of the joint system as $\hat{\rho} \in S(\mathcal{H}^Q\otimes\mathcal{H}^C)$.  Here again we take the closed system unitary evolution to be governed by an arbitrary Hamiltonian $\hat{H}$, which may include both self-Hamiltonians and an interaction Hamiltonian between the $C$ and $Q$ subsystems. We now wish to study whether the above procedure results in a well-defined \textit{classical-quantum} limit – a limit in which the $C$ subsystem may be treated classically, while the generic $Q$ system is still described using standard quantum mechanics. For notational convenience in what follows, we will reserve the use of hats for operators with support on the C system Hilbert space $\mathcal{H}^C$; operators acting on $\mathcal{H}_Q$ alone will be left without.

To adapt the standard classical limit procedure to the case where $C$ is a subsystem, one may use a partial Wigner representation \cite{boucher1988semiclassical,kapral1999mixed}. This provides an equivalent representation of quantum mechanics in which one part of the system is described in terms of a phase space, while the other part remains described by operators in Hilbert space. Specifically, we map operators that act on $\mathcal{H}^Q\otimes\mathcal{H}^C$ to phase-space dependent operators on $\mathcal{H}^Q$ by taking the partial trace with respect to $\hat{A}_{q,p}$ i.e. $\hat{O}\mapsto\tr_C[\hat{A}_{q,p}\hat{O}]$. The only difference from the Wigner representation is that the trace is performed over the $C$ subsystem alone, leaving an operator-valued function of phase space. In this representation, the bipartite quantum state $\hat{\rho}$ is represented by the partial Wigner distribution $\varrho^W(q,p)$, which is an operator-valued function of the phase space associated to the $C$ system, given by
\begin{equation}
    \varrho^W(q,p)=\tr_C [\hat{A}_{q,p} \hat{\rho} ].
\end{equation} By the properties of $\hat{A}_{q,p}$ and $\hat{\rho}$, it follows that $\varrho^W(q,p)$ is a Hermitian-valued operator and is normalised when integrated over phase space and traced over Hilbert space i.e. $\int \tr \varrho^W(q,p)\  dq dp=1$. Analogously to how the real-valued Wigner function is not guaranteed to be positive, the Hermitian operator-valued function $\varrho^W(q,p)$ is not guaranteed to be positive semi-definite for all points in phase space. To study the unitary closed system dynamics of the bipartite quantum system in this representation, one may consider the partial Wigner representation of the Hamiltonian $\hat{H}$, given by the Hermitian operator-valued function of phase space $H^W(q,p)=\tr_C[{\hat{A}_{q,p} \hat{H}}]$. The closed system unitary dynamics then takes the form
\begin{equation} \label{eq: masterL_W}
\begin{split}
    \frac{\partial \varrho^W}{\partial t}=-\frac{i}{\hbar}(H^W\star\varrho^W - \varrho^W\star H^W),
\end{split}    
\end{equation} which is analogous to \eqref{eq: WigEvo} except for the fact that here the quantities are operators that act on $\mathcal{H}^Q$. This may be derived by noting that the relation $\hat{f} \hat{g} \mapsto f\star g$, mapping products of operators to star products of functions, still holds in the case of bipartite operators and their corresponding operator-valued partial Wigner transforms \cite{kapral1999mixed}. This dynamics will appear frequently in what follows, and it is thus convenient to define the associated generator $\mathcal{L}^W$ i.e. the generator of closed system evolution under the Hamiltonian $\hat{H}$ in the partial Wigner representation
\begin{equation} \label{eq: L_W}
\begin{split}
    \mathcal{L}^W=-\frac{i}{\hbar}(H^W\star\dt - \dt\star H^W),
\end{split}    
\end{equation} where here we use $\dt$ to denote the input to the generator.

If the argument was to follow as before, then taking $\hbar\rightarrow 0$ of the closed system unitary dynamics in the partial Wigner representation would lead to one system becoming effectively classical, while the other remaining quantum. Considering equation \eqref{eq: L_W} to $O(1)$ in $\hbar$, the resulting equation 
\begin{equation} \label{eq: CQ_liouville}
\begin{split}
    \frac{\partial \varrho^W}{\partial t}=-\frac{i}{\hbar}[H,\varrho^W]& +\frac{1}{2}(\{H,\varrho^W\} - \{\varrho^W,H\}),
\end{split}
\end{equation} is known as the quantum-classical Liouville equation \cite{kapral1999mixed,Aleksandrov+1981+902+908,gerasimenko1982dynamical}. Here $H$ is the $O(\hbar^0)$ part of $H^W$ and we will refer to this as the classical-quantum Hamiltonian. The first term takes the form of a Liouville von-Neumann term, that describes unitary evolution of density operators. The second term, sometimes referred to as the Alexandrov-Gerasimenko bracket, is a version of the Poisson bracket that is symmetric in the ordering of the phase space dependent operators $H$ and $\varrho^W$.  As before, this form of dynamics will appear repeatedly, and it will be useful to define the corresponding generator
\begin{equation} \label{eq: L_W O(0)}
    \mathcal{L}^W \big\rvert_{O(\hbar^0)} = -\frac{i}{\hbar}[H,\dt] +\frac{1}{2}(\{H, \dt\} - \{\dt,H\}),
\end{equation} which is simply the generator $\mathcal{L}^W$ of \eqref{eq: L_W} truncated to $O(1)$ in $\hbar$.

However, there is a key difference between the Liouville and quantum-classical Liouville dynamics: while the Liouville equation preserves the classicality of the $C$ system, the quantum-classical Liouville equation does not. Although a full discussion of the effective classicality of subsystems will be provided in section \ref{sec: technic}, for the time being we will state that for a quantum state $\hat{\rho}$ on the bipartite Hilbert space $\mathcal{H}^C\otimes \mathcal{H}^Q$ to be effectively classical on the $C$ subsystem, it is necessary for the corresponding operator-valued partial Wigner distibution to be positive semi-definite at all points in phase space i.e
\begin{equation} \label{eq: varrho+}
    \varrho^W(q,p)\succeq 0\quad \forall q,p \in \mathcal{M}.
\end{equation}
Taking \eqref{eq: varrho+} as a necessary condition for effective classicality of the $C$ subsystem guarantees that the state may be written as a positive probability distribution over phase space multiplied by a corresponding quantum state on $\mathcal{H}^Q$ at each point, and is the natural generalisation of \eqref{eq: W+} to operator-valued functions. For a dynamics to preserve the classicality of the $C$ subsystem, it therefore must be the completely-positive on all initial operator-valued functions of phase space. However, while the Liouville equation preserves the positivity of real-valued functions, the quantum-classical Liouville equation does not preserve the positivity of operator-valued functions of phase space \cite{boucher1988semiclassical}. This may be seen by appealing to the recently proved analogue of the Pawula theorem for operator-valued functions \cite{oppenheim2022two} (see also \cite{diosi2023hybrid} for a later discussion of this result). Known as the CQ Pawula theorem, it showed that every trace-preserving, normalised and completely-positive Markovian dynamics on operator-valued functions of phase space is also separated into two classes, with one class truncating at second order in derivatives in phase space, and the other containing an infinite number of higher derivative terms (see Appendix \ref{app: Pawula}). Since the full dynamics of \eqref{eq: L_W} typically truncates at a finite number of derivative terms, the $\hbar \rightarrow 0$ limit helps to bring the resulting form of equations closer to a completely-positive form, by removing all derivative terms second order and higher. However, as we show in Appendix \ref{app: Pawula}, even with these higher order derivatives removed, the resulting dynamics are still not of the required form for complete-positivity.

The problem ultimately lies in the fact that while $\hbar\rightarrow0$ suppresses non-classicality arising from the higher order derivatives in $q$ and $p$, it has no effect on suppressing the entanglement that is generated between the $C$ and $Q$ quantum subsystems. 
Since entanglement may be generated for even linear coupling between subsystems, the quantum-classical Liouville equation \eqref{eq: CQ_liouville} must also generically describe entanglement build up between the $C$ and $Q$ subsystems, and thus the generation of states that are not effectively classical on the $C$ subsystem.

Before moving on to see how one may resolve this, we should first address a technical detail regarding the kinds of Hamiltonian that we consider. Up to this point, we have implicitly assumed that $H^W$, referring to either the Wigner or partial Wigner representation of the Hamiltonian $\hat{H}$, may be written as $H^W= H + O(\hbar^2)$. This assumption holds when $\hat{H}$ is a function of $\hat{q}$ and $\hat{p}$, and in such typical cases, the classical or classical-quantum Hamiltonian $H$ coincides with the function of phase space obtained by making the substitutions $\hat{q}\mapsto q$, $\hat{p}\mapsto p$.  In general however, the Hamiltonian $\hat{H}$ may also depend explicitly on $\hbar$, and in these cases there is no guarantee that $H^W=H+O(\hbar^2)$. However, if $H^W$ contains any terms of $O(\hbar^{-1})$ or higher inverse powers of $\hbar$, there is no well-defined classical limit as $\hbar \rightarrow 0$ \cite{yaffe1982large}, and one may check that the dynamics truncated to $O(\hbar^0)$ in such cases are not positive. The only remaining case is thus where $H^W$ contains $O(\hbar)$ terms. We consider this in Appendix \ref{app: O(hbar)}, and find that it amounts to only a minor modification of the dynamics when $H^W=H+O(\hbar^2)$. For conceptual clarity we therefore present the following analysis under the assumption that $H^W=H+O(\hbar^2)$ – but this, up to a known modification, describes all possible Hamiltonians which permit a classical limit.

\section{Decoherence timescale $\tau$ and a double scaling limit} \label{sec: env}

The preceding section introduced the formalism of the Wigner and partial Wigner representations, and showed how the standard $\hbar \rightarrow 0$ limit is insufficient to describe a classical limit of a subsystem due to the presence of entanglement. In this section, we introduce a simple model of the effect of an environment on the $C$ subsystem, and show how this leads one to a double scaling limit involving the decoherence timescale $\tau$ of this subsystem.

We begin by noting that it is well-understood that $\hbar\rightarrow0$ is not sufficient to ensure classicality, even in single systems. The key observation is that when $\hbar$ is small, but finite, the evolution generated by the Liouville equation will generally map an initial state $W(q,p,t_i)$ in which the higher order terms are negligible to a state at later times $W(q,p,t_f)$ in which they are not \cite{zurek1994decoherence}. The resolution to this problem was to acknowledge that in practice, all quantum systems are open systems, and thus interact with their environments. In this case, the interaction with an environment leads to dispersion in the system, preventing any later states of the Wigner quasiprobability distribution $W(q,p,t_f)$ from becoming overly peaked in phase space and thus preventing the higher order terms contributing, an analysis that has been put on more rigorous footing in recent work \cite{hernandez2023hbar,hernandez2023decoherence}. More generally, acknowledging the role of the environment, which generically acts to decohere the system, has turned out to be an extremely successful way of explaining a number of features in the quantum-to-classical transition \cite{schlosshauer2005decoherence}. 

In what follows, we shall follow the above philosophy by modelling the effect of the environment on the subsystem that will be classicalised. The basic idea is that the interactions with an environment will lead to decoherence on the $C$ subsystem that can break the entanglement with the $Q$ subsystem. In other words, the decoherence induced by an environment will act to replace the quantum correlations between the $C$ and $Q$ subsystems with purely classical correlations, which will ensure that the resulting $\varrho^W$ is positive.

In order to include the effect of an environment, without overly increasing the complexity of the analysis, we will assume that the effective action of the environment is to collapse the $C$ subsystem into a classically definite state. The classically definite states will be taken to be \textit{coherent states}, which are the states with minimum uncertainty in $q$ and $p$. Allowing them to have some squeezing, such that the ratio of the variances in position and momentum is given by $s^2= \frac{\Delta q}{\Delta p}$, we will denote the coherent state with expectation values  $\langle \hat {q} \rangle =q$ and  $\langle \hat {p} \rangle =p$ as $|\alpha_s(q,p)\rangle$ \cite{takahashi1986wigner,gardiner2004quantum}. The environment is then modelled as performing a coherent state POVM with measurement operators $\hat{M}^s_{q,p}=(2\pi\hbar)^{-{1/2}}|\alpha_s(q,p)\rangle\langle \alpha_s(q,p)|$. Assuming for now that the observer has no access to the environmental degrees of freedom, the effect of the environment is a decoherence channel $\hat{\rho} \rightarrow \int dq dp \hat{M}^s_{q,p}\hat{\rho} \hat{M}^s_{q,p}$. In the partial Wigner representation this amounts to a convolution of $\varrho^W$ with a normalised Gaussian with variance $\hbar s^2$ in $q$ and $\hbar s^{-2}$ in $p$. Such a convolution is known as a Weierstrass transform \cite{hirschman2012convolution}, and has the following representation as a differential operator
\begin{equation} \label{eq: env\E vo}
    \mathcal{D} =  e^{ \frac{\hbar s^2}{2} \frac{\partial^2}{\partial q^2} + \frac{\hbar}{2 s^2} \frac{\partial^2}{\partial p^2} } .
\end{equation}
This differential operator $\mathcal{D}$ provides a representation of the decoherence action of an environment, and will prove extremely useful.


Although we have specified the action of the environment as collapsing the states of the system to coherent states, we have not specified over which timescale. To do so, we will specify explicitly that the environment collapses the state of the system over a time $\tau$. This timescale $\tau$ is to be understood to be the decoherence timescale of the $C$ subsystem i.e. the time over which the interaction with the environment has decohered the $C$ subsystem to being in a classically definite state.

The joint specification of the map $\mathcal{D}$ and associated timescale $\tau$ leads to something akin to a trotterised picture of dynamics, in which the effect of the environment is modelled by a decoherence channel that acts at discrete time intervals $\tau$. For now leaving aside the unitary dynamics generated by $\hat{H}$, this explicitly means that the total dynamics in the partial Wigner representation is given by the application of the differential operator $\mathcal{D}$ at times $0,\tau,2\tau,\ldots$ and so on, with no evolution in between. Although different from the standard continuous time dynamics of typical open systems treatments \cite{breuer2002theory}, the advantage of this discrete-time approach is that after each action of the decoherence channel, the state is in a guaranteed classical state.

In order to arrive at a well-defined continuum limit of this discrete-time model of decoherence, one would wish to take the decoherence timescale to zero i.e. $\tau \rightarrow 0$. Since the environment acts to select classical states on the $C$ subsystem, taking this limit ensures that the subsystem is in a classical state at all times. However, since each application of the decoherence map causes $\varrho^W$ to be convolved with a Gaussian with variances proportional to $\hbar$, taking $\tau \rightarrow 0$ while $\hbar$ remains finite would lead to the state of the system becoming infinitely spread in phase space in finite time. To prevent this from occurring, we observe that simultaneously taking the limit that $\hbar\rightarrow 0$ allows, in principle, for an infinite series of convolutions to still give a finite effect on the resulting distribution. To see which rates it may be sensible to take $\hbar$ and $\tau$ to zero, one may consider $t/\tau$ environmental decoherence steps, which gives the overall effect of the environment $\mathcal{G}_t$ after finite time $t$ as
\begin{equation} 
    \mathcal{G}_t=\lim_{\hbar,\tau\rightarrow 0}\mathcal{D}^{t/{\tau}}=\lim_{\hbar,\tau\rightarrow 0}e^{\frac{\hbar}{\tau}\big(\frac{s^2}{2} \frac{\partial^2}{\partial q^2} + \frac{1}{2 s^2} \frac{\partial^2}{\partial p^2}\big)t}.
\end{equation} Remarkably, one can see that when the ratio $\hbar/\tau$ is fixed, the differential operator $\mathcal{G}_t$ corresponds exactly to the semi-group corresponding to a classical diffusion process, with the diffusion rate in $q$ given by $\hbar s^2/\tau$ and diffusion rate in $p$ given by $\hbar s^{-2}/\tau$. Despite starting from a strong measurement at each step in time, the resulting equations of motion describe continuous evolution in time and phase space. This motivates the following double-scaling limit as a candidate method of taking the classical-quantum limit:
\begin{equation} \label{eq: DSCL}
    \hbar\rightarrow 0,\  \tau \rightarrow 0, \quad s.t.\quad  \frac{\hbar}{\tau}=\E, 
\end{equation} where we have chosen $\E $ to denote the constant with dimensions of energy that describes the fixed ratio of the two. As before, the taking of dimensionful quantities to zero should be more carefully interpreted as statements about the relevant scales in the system. Here we may interpret the above double-scaling limit as the statement that the action associated to any observables of interest on the $C$ subsystem are large compared to the scale of $\hbar$, and change over much longer timescales than the decoherence time $\tau$ of the $C$ subsystem. The ratio of the reduced Planck's constant and the decoherence time give a measure of the size of the fluctuations in the system due to the environment, which is captured by the constant $\E $.

Before we move on to study the dynamics that results from taking this limit, it is important to emphasise that the proposed limit should be understood as the application of the double scaling limit to the discrete-time model of decoherence discussed so far. Indeed, it turns out that simply identifying a double scaling limit of the parameter controlling the decoherence timescale and $\hbar$ in a model describing environmental decoherence is not in general sufficient to derive a classical-quantum limit. A straightforward counterexample is provided in Appendix \ref{app: cont}, where we consider a continuous time Lindbladian dynamics with a parameter $\gamma$ controlling the overall decoherence rate in $\hat{q}$ and $\hat{p}$ on the $C$ subsystem. While a double scaling limit $\gamma \rightarrow \infty$, $\hbar \rightarrow 0$ s.t. $\gamma \hbar^2=\E$ leads to identical diffusive evolution in phase space to that described above, the resulting evolution for a bipartite quantum system is not completely-positive, and thus does not describe a well-defined classical-quantum limit. The failure of this approach appears to be related to the lack of equivalence between the two classes of dynamics considered in one of the earliest works on effective classical-quantum dynamics \cite{HalliwelDH}. Although we will not consider this type of set-up any further, it turns out it was the first to appear in the discussion of double scaling classical limits, where the possibility of taking a double scaling limit of $\hbar$ and a measurement rate of a continuous measurement procedure to arrive at diffusive classical evolution on a single system was pointed out in the conclusion of \cite{werner1995classical}. A closer theoretical set-up to ours appears in the context of the quantum Zeno effect \cite{gagen1993continuous}.  More recently, another related model with a different double scaling limit was considered in the context of holography \cite{Kirklin:2020qtv}.

\section{Main results} \label{sec: main}

In this section, we use the discrete-time model of decoherence and associated double scaling limit of the previous section to arrive at the general form of dynamics when one takes the classical-quantum limit we have introduced. This main result is the given in equation \eqref{eq: LcanCQ}. Since some of the technical steps are rather long, we reproduce here only the key conceptual points, and refer the reader to Appendix \ref{app: L} for more details.

In section \ref{sec: env}, the effect of the environment was considered in isolation. However, the key question of interest is to consider how the environment and the free evolution of a generic quantum system interplay in the double-scaling limit we have arrived at. To study this, we must consider the total evolution after a time $\tau$, which should include both the environmental decoherence effects given by $\mathcal{D}$ and the free evolution generated in the partial Wigner representation by $\mathcal{L}^W$. The obvious question then arises of which ordering to choose of the two processes. This point will be made precise in section \ref{sec: quasi} when alternative partial quasiprobability representations are considered, but at this time we will simply postulate a reasonable choice, namely a symmetrised total evolution, in which the action of the environment is divided equally between one part before the free evolution generated by the Hamiltonian, and one part afterwards
\begin{equation} \label{eq: totalW}
    \mathcal{E}^\hbar_\tau=\mathcal{D}^{\frac{1}{2}} e^{\mathcal{L}^W \tau} \mathcal{D}^{\frac{1}{2}} .
\end{equation} The total evolution operator $\mathcal{E}^\hbar_\tau$ describes the action of both the environment and the free evolution on the partial Wigner representation of the bipartite quantum system $CQ$, and we use the superscript and subscript to indicate the functional dependence on both $\tau$ and $\hbar$.

The evolution map $\mathcal{E}^\hbar_\tau$ describes the total change in the partial Wigner representation over a finite decoherence time $\tau$ with a finite value of $\hbar$. In order to take the classical-quantum limit described in \eqref{eq: DSCL} one must consider the infinitesimal evolution in $\tau$ generated when $\tau$ and $\hbar$ are taken to zero such that $\hbar=\E \tau $. To do so, we first set $\hbar=\E \tau$ in $\mathcal{E}^\hbar_\tau$, and consider the generator of the evolution map $\mathcal{E}_\tau:=\mathcal{E}^{\E \tau}_\tau$ in the $\tau\rightarrow 0$ limit, which takes the form
\begin{equation} \label{eq: Ldef}
    \mathcal{L} =  \lim_{\tau \rightarrow 0}\big( \frac{\partial }{\partial \tau } \mathcal{E}_\tau \big) {\mathcal{E}_\tau }^{-1} + \frac{\partial }{\partial \tau }\big(\ln \mathcal{E}_0\big\rvert_{\E =\frac{\hbar}{\tau}}\big).
\end{equation} The first term is the standard form of the generator often used to formally construct time-local dynamics \cite{hall2014canonical,breuer2016colloquium}, and may be found by writing the state of the system at time $\tau$ as $\varrho(\tau)=\mathcal{E}_\tau \varrho(0)$, studying its rate of change in time, and then using the inverse of the evolution map $\mathcal{E}_\tau^{-1}$ to re-express this as a generator as acting on $\varrho(\tau)$. We shall see that this part of the generator captures a large proportion of the dynamics, and importantly the back-reaction of the quantum system on the classical one. However, by construction, this part of the generator only captures the $\tau$-dependent part of the dynamics. In fact, one can check that there is an additional $\tau$-independent component $\mathcal{E}_0$, generated by $-\frac{i}{\E }[H,\ \cdot \ ]$. As discussed in Appendix \ref{app: L}, this term may be accounted for by reintroducing $\hbar=\E \tau$, and computing the generator of this component, corresponding to the second term in \eqref{eq: Ldef}. Since this term only effects the quantum system, the reappearance of $\hbar$ is to be expected: while $\hbar \rightarrow 0$ should be interpreted as the assumption that the relevant classical observables are much larger than $\hbar$, no assumption is made on the scale of relevant quantum observables. From this point on, any appearance of $\hbar$ should be interpreted as characterising the quantum features of the $Q$ subsystem.

To compute the above generator explicitly, one may substitute the evolution map into the form of the generator provided above. Explicitly, by substituting equation \eqref{eq: totalW} into equation \eqref{eq: Ldef} with the $O(\hbar^0)$ expression for $\mathcal{L}^W$ and definition of $\mathcal{D}$ given in equations \eqref{eq: L_W O(0)} and \eqref{eq: env\E vo}, we arrive at
\begin{equation} \label{eq: Lseries}
\begin{split}
    \mathcal{L}=&-\frac{i}{\hbar}[H,\ \cdot \ ]\\
    &+ \frac{1}{2}(1 + e^{\ad{ \frac{-i}{\E } [H,\cdot]}}) \bigg( \frac{\E s^2}{2}\frac{\partial^2 }{\partial q^2}+ \frac{\E }{2s^2}\frac{\partial^2 }{\partial p^2}\bigg) \\
    &+ \frac{e^{\ad{ \frac{-i}{\E } [H,\cdot]}}-1}{\ad{ \frac{-i}{\E } [H,\cdot]}}\bigg( \frac{1}{2}\{H,\cdot\} - \frac{1}{2}\{\cdot,H\}\bigg),
\end{split}
\end{equation} where here the $ad$ denotes the adjoint operation with respect to the generators of the classical-quantum dynamics i.e. $(ad_\mathcal{A} \mathcal{B})\varrho = (\mathcal{A}\mathcal{B}-\mathcal{B}\mathcal{A})\varrho$. Here the first line corresponds to the second term of equation \eqref{eq: Ldef}, while the second and third lines arise from the first term. The complex structure of this part of the generator owes itself to the fact that the generators of the exponential maps that make up $\mathcal{E}_\tau$ do not commute with themselves for all $\tau$. This means that when the derivative in \eqref{eq: Ldef} is taken, one must be careful to use the derivative of the exponential map i.e. if $\mathcal{X}_\tau$ is a classical-quantum generator with dependence on $\tau$, then
\begin{equation} \label{eq: expMap}
    \frac{\partial}{\partial \tau} e^{\mathcal{X}_\tau}=\frac{e^{\ad{\mathcal{X}_\tau}}-1}{\ad{\mathcal{X}_\tau}} \bigg(\frac{\partial \mathcal{X}_\tau}{\partial \tau}\bigg) e^{\mathcal{X}_\tau} ,
\end{equation} which should be understood as a power series of $\ad{\mathcal{X}}$ acting on the derivative of $\mathcal{X}_\tau$, which then acts on the exponential of $\mathcal{X}_\tau$, as is commonly considered in the derivation of the Baker-Campbell-Hausdorff formula \cite{hall2013lie}. 

To further simplify this expression for the generator of dynamics $\mathcal{L}$, one may explicitly compute how the adjoint acts on the phase space derivatives that appear in the various series in \eqref{eq: Lseries}. Amounting to computing part of the Lie algebra corresponding to classical-quantum generators, this allows allows one to map the expressions involving the adjoint action of classical-quantum generators (i.e. $\ad{ \frac{-i}{\E } [H,\ \cdot\ ]}$), to expressions involving the adjoints of quantum operators (i.e. $\ad{ \frac{-i}{\E }H}$). Upon doing so, one arrives at the following form of dynamics 
\begin{equation} \label{eq: LcanCQ}
    \begin{split}
        \mathcal{L}\varrho=&-\frac{i}{\hbar}[ H + H_{eff},\varrho]\quad \quad \quad\\
        &-\frac{1}{2}\frac{\partial}{\partial q}\{L_p^H,\varrho\}_+ + \frac{1}{2}\frac{\partial}{\partial p}\{L_q^H,\varrho\}_+\\
        &+\frac{i s^2}{2}\frac{\partial}{\partial q}[L_q^H,\varrho]+\frac{i}{2s^2}\frac{\partial}{\partial p}[L_p^H,\varrho] \\
        &+\frac{1}{2\E } (\bar{L}\varrho \bar{L}^\dag -\frac{1}{2}\{\bar{L}^\dag\bar{L},\varrho\}_+)\\
        &+ \frac{\E s^2}{2}\frac{\partial^2 \varrho}{\partial q^2}+ \frac{\E }{2s^2}\frac{\partial^2 \varrho}{\partial p^2},
    \end{split}
\end{equation}
where
\begin{equation} \label{eq: Lindblad_def}
\begin{split}
    L_q^H&=\frac{e^{\ad{ \frac{-i}{\E } H}}-1}{\ad{ \frac{-i}{\E } H}}  \bigg(\frac{\partial H}{\partial q}\bigg)\\
    L_p^H&=\frac{e^{\ad{ \frac{-i}{\E } H}}-1}{\ad{ \frac{-i}{\E } H}}  \bigg(\frac{\partial H}{\partial p}\bigg)\\
    \bar{L}&=s L^H_q+ i s^{-1} L^H_p,
\end{split}
\end{equation}
and
\begin{equation} \label{eq: Heff_def}
\begin{split}
    H_{eff}=&\frac{\hbar s^2}{4}\frac{\partial L_q^H}{\partial q}+\frac{\hbar}{4s^2}\frac{\partial L_p^H}{\partial p} \\
    &+\frac{\hbar}{4\E }\sum_{n,m=0}^\infty \frac{C_{nm}}{(n+m+2)!} \\
    &\quad \times \{ \ad{\frac{-i}{\E }H}^n \frac{\partial H}{\partial q},\ad{\frac{-i}{\E }H}^m \frac{\partial H}{\partial p} \}_+ .
\end{split}
\end{equation}
Here $C_{nm}$ denote numerical coefficients given by
\begin{equation}
    C_{nm}= \sum_{r=0}^m\frac{(r+n)!}{r!n!} - \sum_{r=0}^n\frac{(r+m)!}{r!m!},
\end{equation} which we show in Appendix \ref{app: L} are antisymmetric coefficients that may be obtained from the Pascal triangle with integer boundary elements. The operators $L_q^H$ and $L_p^H$ are Hermitian, and have an alternative representation as 
\begin{equation}
    L_z^H=i \E \frac{\partial}{\partial z}\big( e^{-\frac{i}{\E}H}\big) e^{\frac{i}{\E}H},
\end{equation} for $z=q,p$. This can be seen to be equivalent to the definition in \eqref{eq: Lindblad_def} with use of the derivative of the exponential map, as in \eqref{eq: expMap}.

To give some intuition about the dynamics, we sketch the role of each line as follows. The top line describes purely unitary evolution of the quantum system, governed by both the classical-quantum Hamiltonian $H$ and an effective Hamiltonian $H_{eff}$ that depends on $s$ and $\E $. This additional Hamiltonian term arises due to the fluctuations induced by the environment \cite{welton1948some}, and is analogous to the Lamb and Stark shifts that renormalise the bare system Hamiltonian in standard open systems treatments \cite{breuer2002theory}. The second line describes both the free classical evolution and the back-reaction of the quantum system upon it, and we shall see that in Section \ref{sec: two} that this reduces to the symmetrised Poisson bracket appearing in \eqref{eq: CQ_liouville} for a special class of classical-quantum Hamiltonians. The third line describes how random fluctuations in the classical degrees of freedom are correlated with random fluctuations in the unitary dynamics of the quantum system i.e. noisy Hamiltonian quantum dynamics. The fourth line describes the Lindblad portion of the dynamics, which acts to decohere the quantum system into a basis determined by the Lindblad operators $L^H_q$ and $L^H_p$. Finally, the final line describes the previously described diffusion in the classical degrees of freedom, with overall strength proportional to $E_f$ and relative strengths in position and momentum determined by the parameter $s$.

To understand whether the evolution laws given by the above generator are consistent, it is important to check that the dynamics are linear, trace-preserving, and completely-positive on a suitable set of operator-valued functions of phase space. While this seems likely \textit{a priori}, given that the generator above was derived from free evolution and environmental decoherence in a full quantum theory, it is often the case in the study of open quantum systems that approximations lead to violations of one or more of these conditions \cite{breuer2002theory}. In order to check this, we note that the simplified form of the dynamics given in \eqref{eq: LcanCQ} is of the canonical classical-quantum form of dynamics, first written in general form by \cite{oppenheim2022two} (see also \cite{diosi2023hybrid} for a later discussion of this). Any dynamics of this form is linear and trace-preserving, and these properties are straightforward to directly check by hand. In order to check the positivity of a dynamics of this form, one must check a series of positivity conditions given by the CQ Pawula theorem \cite{oppenheim2022two}. The first step is to pick a basis of operators, and phase space degrees of freedom, in which to read off certain decoherence,  back-reaction and diffusion matrices. Picking the basis $(q,p), (L_q^H,L_p^H)$ for convenience, one may refer to the general form given in Appendix \ref{app: L} to see that that the decoherence $D_0$, back-reaction $D_1$ and diffusion $D_2$ are given by
\begin{equation} \label{eq: D0}
\begin{split}
D_0&=\begin{pmatrix}
    \dfrac{s^2}{2\E } & -\dfrac{i}{2\E }\\[2ex]
    \dfrac{i}{2\E } & \dfrac{1}{2\E s^2}    
    \end{pmatrix}  \\
\end{split}
\end{equation}
\begin{equation}\label{eq: D1}
\begin{split}
D_1&=\begin{pmatrix}
    \dfrac{is^2}{2} & \dfrac{1}{2}\\[2ex]
    -\dfrac{1}{2} & \dfrac{i}{2s^2}    
    \end{pmatrix} \\
\end{split}
\end{equation}
\begin{equation}\label{eq: D2}
\begin{split}
D_2&=\begin{pmatrix}
    \E s^2 & 0\\[2ex]
    0 & \dfrac{\E }{s^2}    
    \end{pmatrix}. \\
\end{split}
\end{equation}
The most basic requirements for positivity in the CQ Pawula theorem are the same as those of the Lindblad and Fokker-Planck equations. For the quantum degrees of freedom, these are the requirements that for all points in phase space the total Hamiltonian $H+H_{eff}$ is Hermitian and the decoherence matrix $D_0$ is positive semi-definite. For the classical degrees of freedom, it is that the real matrix $D_2$ is positive semi-definite for all points in phase space. One may check that these properties do indeed hold, with $H_{eff}=H_{eff}^\dag$ following from $H=H^\dag$. The key result of the classical-quantum Pawula theorem, is that the remaining conditions sufficient for complete-positivity of a classical-quantum dynamics are that  $(\mathbb{I}-D_2D_2^{-1})D_1=0$ and that $D_0\succeq D_1^\dag D_2^{-1} D_1$, where $\mathbb{I}$ denotes the identity matrix of the dimension of the classical degrees of freedom, and $D_2^{-1}$ denotes the pseudoinverse of $D_2$. The first condition ensures that any classical degree of freedom that experiences quantum back-reaction has noise in it, and this holds here since $D_2$ is full-rank. The second condition, known as the decoherence-diffusion trade-off \cite{decodiff2}, ensures that decoherence in the quantum system is sufficiently large to to be compatible with the rate of information gain about it by the classical system. One may explicitly check this condition with the above matrices and see that the decoherence-diffusion trade-off is satisfied, and in fact is saturated as $D_0= D_1^\dag D_2^{-1} D_1$. 
Thus, analogously to the standard classical limit of the Wigner distribution, the classical-quantum limit presented here arrives at a dynamics that is positive on all initial operator-valued functions of phase space.

As a final technical remark, it is worth noting that the above form of dynamics may be straightforwardly generalised from one describing a single pair of phase space coordinates $(q,p)$ to one describing many i.e. $(q_1,\ldots,q_n,p_1,\ldots,p_n)$. To see this, we note that generalising the $C$ subsystem to $[\hat{q}_i,\hat{p}_j]=i\hbar \delta_{ij}$ for $i,j=1,\ldots, n$ and assuming decoherence into a tensor product of coherent states on each subsystem, each described by a parameter $s_i$, changes the decoherence map $\mathcal{D}$ and the partial Wigner evolution $\mathcal{L}^W$ in a particularly simple way. In particular, the only change in $\mathcal{D}$ is to include a sum from $i=1$ to $i=n$ in the exponential, while $\mathcal{L}^W$ to $O(\hbar^0)$ has the inclusion of the same sum in the Poisson bracket term. Since these lead to the second and third lines in equation \eqref{eq: Lseries} respectively, it follows by linearity that the only change in the final result of equation \eqref{eq: LcanCQ} is the inclusion of a sum over the pairs of phase space degrees of freedom i.e. performing the substitution $s\mapsto s_i$, $q\mapsto q_i$, $p\mapsto p_i$ and summing from $i=1,\ldots,n$ over any terms that these indices explicitly appear in. This form of dynamics leads to $D_0$, $D_1$ and $D_2$ matrices that are block diagonal, with each block taking the same form as those given in \eqref{eq: D0}, \eqref{eq: D1} and \eqref{eq: D2}, and thus the resulting dynamics is completely-positive as before.

\section{Effective classical-quantum states and subset positivity} \label{sec: technic}

The analysis of both the insufficiency of the $\hbar\rightarrow 0$ limit and the apparent success of the double scaling limit have thus far been presented using the partial Wigner quasiprobability distribution $\varrho^W$. However, the positive semi-definiteness of $\varrho^W$ was stated to only be a \textit{necessary} condition for the classicality of a subsystem. In this section, we introduce a general definition of partial quasiprobability representations, and the effective classicality of a subsystem, show that the positivity of a partial Glauber-Sudarshan quasiprobability distribution $\varrho^P$ provides both necessary \textit{and sufficient} conditions for the $C$ subsystem to be effectively classical. The considerations in this section and the next do not change the main result of equation \eqref{eq: LcanCQ}, and those interested in understanding this general form of classical-quantum dynamics better may instead choose to go straight to section \ref{sec: traj} or \ref{sec: two}.

We start by defining a notion of effective classicality for a single quantum system. To motivate this, it will be useful to introduce two common alternatives to the Wigner quasiprobability representation, the $Q$ and $P$ representations, well-known for their use in quantum optics \cite{husimi1940some,sudarshan1963equivalence,glauber1963coherent}. In the $Q$ representation, states are represented by the Husimi distribution $Q$, which is defined explicitly as
\begin{equation} \label{eq: Q}
    Q(q,p)= \tr\bigg[\frac{|\alpha_s(q,p)\rangle \langle \alpha_s(q,p)|}{2\pi\hbar} \hat{\rho}  \bigg],
\end{equation}
while in the $P$ representation, states are represented by the Glauber-Sudarshan distribution $P$, which is defined implicitly by
\begin{equation} \label{eq: P}
    \hat{\rho}=\int P(q,p) |\alpha_s(q,p)\rangle \langle \alpha_s(q,p)| dq dp.
\end{equation} While the $Q$ distribution is always a well-defined function of phase-space, the $P$ distribution in general must be understood as a generalised function \cite{klauder1965diagonal,gardiner2004quantum}. These expressions define the representation of states in each respective quasiprobability theory, and implicitly define the representation of measurements (see Appendix \ref{app: effClass}).

One may identify the effective classicality of a single quantum system by studying the positivity of one of these representations. By construction, the Husimi $Q$ distribution is positive for all quantum states. Therefore, the positivity of the $Q$ distribution cannot be useful as a measure of classicality, since under this definition all quantum states, even those in non-local superpositions, would be understood as effectively classical. On the other hand, one sees that the positivity of the $P$ distribution has a clear physical interpretation –  that the quantum state can be written as a statistical mixture of coherent states i.e. a classical mixture of ``the most classical states". For this reason, we take the positivity of the $P$ representation to be a sufficient and necessary condition for the effective classicality of a single quantum system, as is common within quantum optics \cite{mandel1986non,ferrie2011quasi}. Since the $W$ representation is related to the $P$ representation by a Weierstrass transform (see Appendix \ref{app: W,P,Q}), if the $P$ distribution is positive, the Wigner distribution will also be positive. This explains the original statement in section \ref{sec: wig} that the positivity of the Wigner function is a necessary condition for the classicality of a single quantum system.

Although the above discussion allows one to identify a definition of effective classicality, it glosses over the fact that in general, the effective classicality of a given quantum state should be understood as being with respect to a given set of measurements. The general framework for characterising when a set of quantum states and measurements permit a classical explanation was first provided in \cite{spekkens2008negativity}. We provide a summary of this framework in Appendix \ref{app: effClass}, and show that using it, one can understand the positivity of the $P$ distribution to guarantee that the statistics of any measurement made on the quantum system can be explained using an entirely classical model.

We will now generalise the above discussion to study the effective classicality of subsystems. To make this concrete, we begin by defining in general terms the notion of a {\it partial quasiprobability representation}. Recall that a more general treatment of measurements is that of POVMs $\{\hat{E}_i\}$, where $\hat{E}_i$ denote the POVM elements. A partial quasiprobability representation $R$ is the assignment to every state $\hat{\rho}$ and every set of POVM elements $\hat{E}_i$ the operator-valued functions of phase space $\varrho^R$ and $E^R_i$ acting on $\mathcal{H}^Q$, in such a way that
\begin{equation} \label{eq: partialRep}
    \tr[\hat{\rho}\hat{E_i}]=\int dz\  \tr[\varrho^R (z) E^R_i(z)] .
\end{equation}
Here the trace on the left-hand side is over the $C$ and $Q$ subsystem Hilbert spaces, while the trace on the right-hand side is just over $\mathcal{H}^Q$. By definition, every measurement may be represented in this way, and thus the partial quasiprobability representation provides an entirely equivalent description of bipartite quantum mechanics. The partial Wigner representation described in Section \ref{sec: wig} provides an example of this. Note that in this example the same map is applied to both states $\hat{\rho}$ and POVM elements $\hat{E}_i$ to generate the representation, but in general the states and observables are treated differently.

To identify when a given set of bipartite quantum states $\{\hat{\rho}_\lambda\}$ and measurements $\{\{\hat{E}_i\},\{\hat{F}_i\},\ldots\}$ may be described using an effectively classical subsystem, it is necessary to study the positivity of their representations. This was first demonstrated in \cite{spekkens2008negativity}, where the criterion for whether a given set of quantum states and measurements could be modelled classically was identified as when the representations of both the states and POVM elements were non-negative real-valued functions of phase space. To generalise to the case of an effectively classical subsystem, we will say that {\it a set of states and POVMs admit an effective classical-quantum description} whenever there exists a representation $R$ in which $\varrho^R$ and $E^R_i$ are positive semi-definite for all $z$ in phase space, by direct analogue with the purely classical case. As in the case of defining an effective classical description of a quantum system \cite{spekkens2008negativity,bartlett2012reconstruction}, only a restricted set of all measurements and states in quantum theory permit an effective classical-quantum description.

For a restricted set of measurements, many quantum states may admit an effective classical-quantum description of the combined set of measurements and states. However, a special class of states are those which may be modelled using a classical-quantum description for all possible bipartite measurements on the system. Translated to the technical language above, we will call a bipartite density operator $\hat{\rho}$ an {\it effective classical-quantum state} whenever there exists a representation where the corresponding partial quasiprobability distribution is positive semi-definite $\varrho^R \succeq 0$ and the representation of all POVMs are positive semi-definite. This provides an operationally relevant definition of states with an effective classical subsystem, since it means that regardless of the form of measurement performed on the joint bipartite quantum system, the statistics are reproducible via an underlying classical-quantum (or partially non-contextual) model \cite{spekkens2008negativity}. 

A second concept that is important to introduce is that of {\it subset-positivity}. In Section \ref{sec: wig}, the notion of positivity of dynamics was introduced, and used to argue for the validity of the Liouville equation as classical equation of motion, and against the quantum-classical Liouville equation as having describing a genuinely classical subsystem. The key property is that the positivity of the dynamics was considered on the set of all positive semi-definite operator-valued functions, that we will denote $\mathbf{S}$. However, there also exist dynamics which although they do not preserve the positivity of all initial real or operator-valued functions of phase-space states, do preserve positivity of on a subset of initial conditions. For a given subset of all positive semi-definite functions $\mathbf{\Lambda} \subset \mathbf{S}$, we state that a dynamics is {\it $\mathbf{\Lambda}$-positive} if it is positive for all initial states belonging to that subset. Since subset-positive dynamics need not positive on all initial states, it also need not be characterised by the Pawula \cite{pawula1967generalizations} or CQ Pawula \cite{oppenheim2022two} theorems described in Appendix \ref{app: Pawula}.

To illustrate these two notions, we consider two important examples of partial quasiprobability representations, derived from the $Q$ and $P$ representations previously introduced. In particular, we may define the partial Husimi distribution $\varrho^Q$ explicitly via
\begin{equation}
    \varrho^Q(q,p)= \tr_C\bigg[\frac{|\alpha_s(q,p)\rangle \langle \alpha_s(q,p)|}{2\pi\hbar} \hat{\rho}  \bigg],
\end{equation}
and the partial Glauber-Sudarshan distribution $\varrho^P$ implicitly by
\begin{equation} \label{eq: varrhoP}
    \hat{\rho}=\int \varrho^P(q,p) \otimes |\alpha_s(q,p)\rangle \langle \alpha_s(q,p)| dq dp.
\end{equation} Like $\varrho^W$, both $\varrho^Q$ and $\varrho^P$ are normalised to 1 when traced over Hilbert space and intergrated over phase space, and are useful for illustrating different properties of a given bipartite quantum state $\hat{\rho}$. The partial Husimi $\varrho^Q$ is an operationally relevant quantity, giving the actual probabilities and corresponding quantum states on $Q$ of a coherent state POVM with measurement operators $\hat{M}_{q,p}$ on the $C$ subsystem, and is consequently positive semi-definite for all $q,p$. A consequence of the non-orthogonality of the coherent states is that the set of all operator-valued functions $\varrho^Q$ form a strict subset $\mathbf{H}\subset \mathbf{S}$ of positive operator-valued functions, in particular not including those with uncertainty in position and momentum less than the Heisenberg bound \cite{takahashi1986wigner,takahashi1989distribution,diosi2000quantum}. By contrast, the partial Glauber-Sudarshan $\varrho^P$ is not necessarily positive semi-definite at all points in phase space \cite{gardiner2004quantum}, but when it is, one may see from its definition that the bipartite quantum state is separable between the classical and quantum subsystems i.e. contains no entanglement \cite{nielsen2002quantum}. 

In addition to guaranteeing that there is no entanglement between the $C$ and $Q$ subsystems, the positive semi-definiteness of the partial Glauber-Sudarshan distribution turns out to provide sufficient and necessary conditions for the underlying bipartite quantum state to be an effective classical-quantum state. To see this, we may substitute the definition of $\varrho^P$ given in equation \eqref{eq: varrhoP} into equation \eqref{eq: partialRep} to see that the representation of POVM elements in the partial $P$ representation are in fact given by the partial $Q$ representation, and thus are always positive semi-definite. By the definition given above, if the partial Glauber-Sudarshan representation $\varrho^P$ for a bipartite state $\hat{\rho}$ is positive, this state  must therefore be an effective classical-quantum state. The positivity of $\varrho^P$ therefore provides sufficient and necessary conditions for the underlying bipartite quantum system to have an effectively classical $C$ subsystem. Since the partial Wigner $\varrho^W$ is related to $\varrho^P$ by a Weierstrass transform (see Appendix \ref{app: W,P,Q}), any positive semi-definite $\varrho^P$ necessarily implies that $\varrho^W$ is also positive, justifying the original claim that $\varrho^W \succeq 0$ is a neccessary condition for an effective classical subsystem.

For the other notion introduced, we note that unitary quantum dynamics in the partial Husimi representation provides an example of subset-positive dynamics. Since the partial Husimi representation is always positive, the unitary dynamics in Hilbert space induces a positive dynamics on partial Husimi distributions \cite{diosi2000quantum}. However, this map is not positive on all initial states, but instead is $\mathbf{H}$-positive. While this subset-positive dynamics has many interesting features, the positivity should not be conflated with the interpretation as having an effectively classical subsystem, for the reason that all dynamics, even those that generate large amounts of entanglement, may be represented in this way.

We conclude this discussion by highlighting two subtleties of studying effectively classical subsystems. Firstly, it is important to emphasise that a subsystem that is not effectively classical may become effectively classical when the $Q$ subsystem is traced out. In particular, it is possible to find examples of $\varrho^P$ that are not positive despite the fact that their trace $\tr\varrho^P$ is. As opposed to being a failure of the definitions we provide, this captures an important feature of classicality: namely that a given system is only classical with respect to a set of other systems that are accessible to measure. Indeed, the $C$ system will in general be expected to be entangled with another unobserved system such as an environment \footnote{In particular, the suppression of entanglement between the $C$ and $Q$ subsystems in the set-up we consider must result from entanglement generation between $C$ and its external environment. We thank an anonymous reviewer for pointing this out.}, or have entanglement amongst unobserved constituent parts (e.g. as in \cite{Kirklin:2020qtv}). Secondly, once one goes beyond a single quantum subsystem that is being classicalised, it is not necessarily natural for the positivity of the partial Wigner function to even be a necessary condition for classicality. This is due to the possibility of entanglement amongst the subsystems that make up the $C$ subsystem, which nevertheless is not manifest at a macroscopic level. Although beyond the scope of the current work, in this case one expects that some entanglement should be permitted amongst the individual subsystems that make up the classical system, and thus the allowed set of measurements should be restricted to reflect that this entanglement is not detectable.

\section{Equivalence between partial quasiprobability representations} \label{sec: quasi}

In this section, we shed some light onto why the dynamics of equation \eqref{eq: LcanCQ} is completely-positive on all operator-valued functions of phase space, and on the original choice of operator ordering in the definition of $\mathcal{E}^\hbar_\tau$, by studying the dynamics of the partial Husimi $\varrho^Q$ and partial Glauber-Sudarshan $\varrho^P$ distributions introduced in the previous section. In doing so, we demonstrate that the classical-quantum limit we have arrived at preserves the effective classicality of the $C$ subsystem.

To study the total dynamics of the partial Glauber-Sudarshan and partial Husimi distributions in the classical-quantum limit, we first note that the decoherence channel used to model the environment in these representations is identical to that of the partial Wigner distribution, and so may be modelled as before as $\mathcal{D}$. To study the unitary dynamics generated by the Hamiltonian in each representation, we show in Appendix \ref{app: W,P,Q} how one may find the generators of the partial Husimi $\mathcal{L}^Q$ and the partial Glauber-Sudarshan $\mathcal{L}^P$ by mapping first to the Wigner distribution by the differential operator $\mathcal{D}^{\mp\frac{1}{2}}$, using the free Wigner evolution, and then mapping back using the inverse $\mathcal{D}^{\pm\frac{1}{2}}$, for $\varrho^Q$ and $\varrho^P$ respectively. Considering the corresponding generators to $O(1)$ in $\hbar$, we find the following generator of partial Husimi evolution
\begin{equation} \label{eq: L_Q}
\begin{split}
    \mathcal{L}^Q\big\rvert_{O(\hbar^0)}=&-\frac{i}{\hbar}[H,\dt] +\frac{1}{2}(\{H, \dt\} - \{\dt,H\}) \\
    &-\frac{is^2}{2}[\frac{\partial H}{\partial q},\frac{\partial \dt}{\partial q}] -\frac{i}{2s^2}[\frac{\partial H}{\partial p},\frac{\partial \dt}{\partial p}] \\
    &-i[\frac{s^2}{4}\frac{\partial^2 H}{\partial q^2}+\frac{1}{4s^2}\frac{\partial^2 H}{\partial p^2},\dt],
\end{split}
\end{equation} which was first written down in \cite{diosi2000quantum}, though without the final term, and the following generator of partial Glauber-Sudarshan evolution
\begin{equation} \label{eq: L_P}
\begin{split}
    \mathcal{L}^P\big\rvert_{O(\hbar^0)}=&-\frac{i}{\hbar}[H,\dt] +\frac{1}{2}(\{H, \dt\} - \{\dt,H\}) \\
    &+\frac{is^2}{2}[\frac{\partial H}{\partial q},\frac{\partial \dt}{\partial q}] +\frac{i}{2s^2}[\frac{\partial H}{\partial p},\frac{\partial \dt}{\partial p}] \\
    &+i[\frac{s^2}{4}\frac{\partial^2 H}{\partial q^2}+\frac{1}{4s^2}\frac{\partial^2 H}{\partial p^2},\dt].
\end{split}
\end{equation} Using these, one may then construct the generator of evolution $\mathcal{E}^\hbar_\tau$ as in \eqref{eq: totalW} and take the double-scaling limit as described previously to find the generator of the dynamics. However, in order to derive the same evolution map, and thus the same generator, one can check that one must choose different operator orderings depending on the representation! In particular, one can see from the above argument using $\mathcal{D}^{\pm \frac{1}{2}}$ to map between representations, that three distinct operator orderings of the free evolution and the environmental decoherence steps lead to the same evolution map:
\begin{equation}
    \mathcal{E}^\hbar_\tau=\ e^{\mathcal{L}^Q \tau}\mathcal{D}=\mathcal{D}^{\frac{1}{2}}\ e^{\mathcal{L}^W \tau}\mathcal{D}^{\frac{1}{2}}=\mathcal{D}e^{\mathcal{L}^P \tau} .
\end{equation} The key observation to understand the difference in operator ordering in each case is to note that the environment plays a different role in each partial quasiprobability representation in order to maintain classicality. As discussed in section \ref{sec: technic}, the unitary dynamics in the partial Husimi representation are only positive on initial states with sufficient spread in phase space. Consequently, in this representation the decohering action of the environment must be taken \textit{before} the unitary evolution, such that any arbitrarily peaked states in phase space are first convoluted before they are evolved. Conversely, in the partial Glauber-Sudarshan representation, the state $\varrho^P$ is only positive when all entanglement has been removed; in this case the environment acts \textit{after} the unitary evolution to ensure any entanglement built up by the unitary evolution is destroyed at the end of each step. Since the partial Wigner representation $\varrho^W$ lies exactly half-way between $\varrho^Q$ and $\varrho^P$ by Wierstrass transform (see Appendix \ref{app: W,P,Q} for more details), the original symmetrised dynamics postulated in \eqref{eq: totalW} turns out to be exactly that which performs both steps in half-measure. In all of these cases, the map that is defined is completely-positive on all positive semi-definite operator-valued functions.

The above analysis also guarantees that the dynamics of equation \eqref{eq: LcanCQ} preserves the effective classicality of the $C$ subsystem. As discussed in Section \ref{sec: technic}, the positivity of the partial Glauber-Sudarshan probability distribution provides sufficient and necessary conditions for the quantum state of the bipartite system to be an effective classical-quantum state. Thus, by here explicitly showing that the dynamics of $\varrho^P$ are also positive, we guarantee that the $C$ subsystem may be treated as effectively classical in the double scaling limit. This may be equivalently argued using the fact that the map between the different representations becomes the identity in the limit that $\hbar \rightarrow 0$, and thus that $\varrho^W$ coincides with $\varrho^P$ in the classical limit. For the same reason, ensuring that the dynamics in the three representations all agree, as it does above, provides an important consistency check on the validity of any classical-quantum dynamics arising from a classical limit.

\section{Trajectories in the classical-quantum limit} \label{sec: traj}

We assumed up to this point that the observer has no access to the environmental degrees of freedom that store the information about the $C$ subsystem. However, one could assume that the observer has sufficient information about the environment to reconstruct the outcome of the effective coherent state POVM that it induces at each time step \cite{blume2005simple,zurek2009quantum,wiseman_milburn_2009}. In this case, the observer has access to the classical system's trajectory in phase space, and their best estimate of the quantum state deduced from the motion of the classical system leads to a quantum trajectory in Hilbert space. 

The general form of dynamics describing classical-quantum trajectories was first given in \cite{layton2022healthier} (see also \cite{diosi2023hybrid} for a later discussion of these points). This work allows for completely-positive classical-quantum master equations, of which Equation \eqref{eq: LcanCQ} is a special case, to be unravelled into coupled stochastic trajectories of the classical phase space degrees of freedom $Z_t$ and mixed quantum states $\rho_t$ conditioned on these classical trajectories. A key result of \cite{layton2022healthier} was that when the trade-off is saturated in the form $D_0=D_1^\dag D_2^{-1} D_1$, any initially pure state of the quantum system remains pure conditioned on the classical trajectory. Since this is the case here, this means that the general classical-limit dynamics of Equation \eqref{eq: LcanCQ} may be unravelled in terms of pure quantum states $|\psi\rangle_t$ which are unique for a given classical trajectory. Defining a column vector $Z_t=(q_t,p_t)^T$ for the classical degrees of freedom, and the operator-valued column vector $L=(L_q^H,L_p^H)^T$, the stochastic dynamics takes the form
\begin{equation} \label{eq: unrav_C}
    \begin{split}
        dZ_t=(D_1^*+D_1)\langle L \rangle dt + \sigma dW_t,
    \end{split}
\end{equation}
\begin{equation}
    \begin{split} \label{eq: unrav_Q}
        d|\psi\rangle_t=&-\frac{i}{\hbar} (H + H_{eff})|\psi\rangle_t dt\\
        &+  (L-\langle L \rangle)^T D_1^\dag \sigma^{-T}  |\psi\rangle_t dW_t \\
        &-\frac{1}{2}(L-\langle L \rangle)^T (D_0^T L - D_0 \langle L \rangle )|\psi\rangle_t dt.
    \end{split}
\end{equation}
Here $\langle L \rangle=\langle \psi|_t L |\psi\rangle_t$, $\sigma$ denotes any $2\times2$ matrix such that $\sigma \sigma^T = D_2$, and $dW_t = (dW^1_t, dW^2_t)^T$ denotes a column vector of two uncorrelated Wiener processes. The equations are formally identical to those used to describe continuous quantum measurement and the associated measurement signals, and we refer the reader to \cite{2006Jacob} for an excellent introduction to this formalism.

The unravelled form of dynamics makes clear that the semi-classical limit we present here does not lead to any loss of quantum information, provided an observer has access to the full classical trajectory \cite{layton2022healthier}. Since this originates from a full quantum theory, we see that in principle the irreversibility introduced by tracing out an environment may be partially recovered in the classical limit. 

The most important practical use of the stochastic unravelling equations we provide here is for the simulation of dynamics in the classical-quantum limit. For the same reasons that unravellings of Lindbladian dynamics provide an important tool for studying open quantum systems \cite{breuer2002theory}, unravellings of classical-quantum master equations allow one to study both the mean values and statistics of various classical and quantum observables. As discussed for the master equation approach, these equations may also be generalised to the case where there are $n$ classical degrees of freedom $(q_1,\ldots, q_n,p_1,\ldots,p_n)$: in this case one must take $Z_t=(q_{1,t},\ldots, q_{n,t},p_{1,t},\ldots,p_{n,t})^T$, $L=(L_{q_1}^H,L_{p_1}^H,\ldots,L_{q_n}^H,L_{p_n}^H)^T$, $dW_t = (dW^1_t, \ldots, dW^{2n}_t)^T$, and take $D_0$, $D_1$ and $D_2$ to be block diagonal, with each block given by the form in equations \eqref{eq: D0}-\eqref{eq: D2}, as discussed in section \ref{sec: main}. With these modifications to the above equations, the dynamics of a wide variety of many body systems with bosonic subsystems may be studied numerically in the classical-quantum limit, and we refer the reader to \cite{layton2022healthier} for some basic examples of the simulation of classical-quantum unravellings.

\section{Two special cases of dynamics} \label{sec: two}

The general form of generator, given in equation \ref{eq: LcanCQ} and their corresponding unravellings in \eqref{eq: unrav_C} and \eqref{eq: unrav_Q} are the main results from this work, describing the general form of dynamics for a bipartite Hamiltonian $\hat{H}$ in the double-scaling classical limit on one subsystem. To gain some more insight into what this dynamics predicts, we will consider now two special cases.

The first case we will consider is the effect of the double-scaled classical limit on a single system. To study this, one can take a bipartite quantum Hamiltonian of the form $\hat{H}=(\hat{p}^2/2m + V(\hat{q}) )\otimes \mathds{1}$, where $\mathds{1}$ is the identity operator on the $Q$ subsystem i.e. a Hamiltonian with trivial action on the $Q$ subsystem. The corresponding classical-quantum Hamiltonian may be computed to be $H=(p^2/2m+V(q)) \mathds{1}$, and defines the operators $L_p^H=(p/m) \mathds{1}$, $L_q^H=\partial_q V(q)\mathds{1}$, $H_{eff}=0$. Using these definitions in the general dynamics \eqref{eq: LcanCQ} one finds that the unitary, Lindbladian, and mixed derivative-commutator terms all vanish, and the mixed derivative-anticommutator terms combine to give the Poisson bracket. This gives the following stochastic dynamics on the classical system
\begin{equation} \label{eq: L_C}
    \frac{\partial \varrho}{\partial t}=\{H,\varrho\} + \frac{\E s^2}{2}\frac{\partial^2 \varrho}{\partial q^2}+ \frac{\E }{2s^2}\frac{\partial^2 \varrho}{\partial p^2},
\end{equation} which describes diffusion around the classical Liouville equation given in \eqref{eq: C_liouville}. This example shows that the idea that that the limit is specific to subsystems is not neccesary – rather the double scaling limit we find provides a general notion of a ``stochastic classical limit", that happens to also give consistent evolution when it is applied to subsystems alone. Although the existence of stochastic classical limits are somewhat of a folk wisdom in physics, the earliest concrete proposal we have found in the literature is a discussion in \cite{werner1995classical}.

A second limiting case of the above dynamics is to consider the dynamics under the approximation 
\begin{equation}
\ad{\frac{-i}{\E }H}^n \bigg(\frac{\partial H}{\partial z}\bigg)\approx 0,
\end{equation} for $z=(q,p)$ and $n>0$. This is true exactly when $H(q,p)$ is self-commuting i.e. when $[H(z),H(z^\prime)]=0$ for all $z,z^\prime$ in phase space, and has an error of $O(\hbar^n)$ when the classical-quantum Hamiltonian takes the form $H=(p^2/2m_C) \mathds{1}+ P^2/2m_Q+V(q\mathds{1}-Q)$, where $Q$ and $P$ are operators on the quantum subsystem satisfying $[Q,P]=i\hbar$. Making this approximation, we find that $L^H_q=\partial_q H$, $L^H_p=\partial_p H $ and $H_{eff}=(\hbar s^2/4)\partial_q^2 H   +(\hbar /4s^2)\partial_p^2 H$.  The dynamics in \eqref{eq: LcanCQ} then reduces in form to the following
\begin{equation} \label{eq: Lapprox}
    \begin{split}
        \frac{\partial \varrho}{\partial t}=&-\frac{i}{\hbar}[ H,\varrho] +\frac{1}{2}\big(\{H,\varrho\} - \{\varrho,H\}\big) \\
        &+\frac{is^2}{2}[\frac{\partial H}{\partial q},\frac{\partial \varrho}{\partial q}] +\frac{i}{2s^2}[\frac{\partial H}{\partial p},\frac{\partial \varrho}{\partial p}] \\
        &+i[\frac{s^2}{4}\frac{\partial^2 H}{\partial q^2}+\frac{1}{4s^2}\frac{\partial^2 H}{\partial p^2},\varrho] \\
        &+\frac{1}{2\E } (\bar{L}\varrho \bar{L}^\dag -\frac{1}{2}\{\bar{L}^\dag\bar{L},\varrho\}_+) \\
        &+ \frac{\E s^2}{2}\frac{\partial^2 \varrho}{\partial q^2}+ \frac{\E }{2s^2}\frac{\partial^2 \varrho}{\partial p^2},
    \end{split}
\end{equation} where we have defined the Lindblad operator $\bar{L}=s L^H_q+ i s^{-1} L^H_p$ as in equation \eqref{eq: Lindblad_def}. The first line gives the unitary evolution and Alexandrov bracket from the quantum-classical Liouville equation \eqref{eq: CQ_liouville}. However, the second and third lines, formed from $H_{eff}$ and the mixed derivative-commutator terms, contain exactly the additional terms associated to the dynamics of the partial Glauber-Sudarshan representation i.e. the first three lines give $\mathcal{L}^P\rvert_{O(\hbar^0)}$, previously found in \eqref{eq: L_P}. We thus see that the total dynamics is exactly the dynamics of the partial Glauber-Sudarshan representation to lowest order in $\hbar$, with additional terms corresponding to noise in the classical and quantum systems. Since the approximation made above occurs at the level of the operators, the complete positivity of the dynamics is unchanged, and thus may still be unravelled by using the simplified forms of the operators $L_q^H$, $L_p^H$ and $H_{eff}$ in equations \eqref{eq: unrav_C} and \eqref{eq: unrav_Q}.

The majority of work in the literature on completely-positive classical-quantum dynamics, including earlier work by the present authors, concluded that the natural form of dynamics would take the form of the quantum-classical Liouville equation with minimal additional noise terms to ensure positivity \cite{diosi1995quantum,oppenheim2023postquantum, layton2022healthier}. However, as the above example shows, when derived in a physical manner from a full quantum theory, a more natural model is instead the $O(\hbar^0)$ partial Glauber-Sudarshan dynamics of \eqref{eq: L_P} supplemented with the minimal terms necessary for positivity. This result seems particularly reasonable when one considers that it is the positivity in the partial Glauber-Sudarshan distribution, and not the partial Wigner distribution, that guarantees the classicality of the $C$ subsystem, as discussed in Section \ref{sec: quasi}. 

\section{The $\E \rightarrow 0$ limit} \label{sec: E}

The double scaling limit we have presented leads generically to irreversible dynamics, with the parameter characterising the diffusion in the classical system given by $\E $. A question we now turn to is whether one may recover a deterministic evolution, as in the standard classical limit, by tuning this free parameter.

The first example to look at is the result of the double scaling classical limit on a single system. The dynamics in this case was computed earlier in equation \eqref{eq: L_C}, taking the form of Hamiltonian dynamics with additional diffusion in both position and momentum proportional to $\E $. In the limit $\E \rightarrow 0$, one thus recovers the Liouville equation \eqref{eq: C_liouville}, i.e. deterministic evolution under the classical Hamiltonian. This additional $\E \rightarrow 0$ limit may be physically interpreted as saying that if one considers large macroscopic scales, any noise due to the environment is negligible, and thus reversible Hamiltonian dynamics is recovered. 
Since the Liouville equation \eqref{eq: C_liouville} was previously obtained directly from the standard $\hbar \rightarrow 0$ limit, we see that when applied to single systems, the stochastic notion of a classical limit that we have presented reduces to the standard notion in the $\E \rightarrow 0$ limit.

Given that the $\E \rightarrow 0$ limit recovers a deterministic classical limit on a single system, it is interesting to consider whether the same may be true when one considers the classical limit of a subsystem. To explore this question, we will first consider the limiting case described in equation \eqref{eq: Lapprox} for self-commuting classical-quantum Hamiltonians. In this case, the parameter appears in two places: proportional to the the strength of classical diffusion, and inversely proportional to the strength of the decoherence on the quantum system. One thus sees that while taking $\E $ to be small reduces the amount of classical diffusion, it leads to very large decoherence on the quantum degrees of freedom in a basis determined by the Lindblad operator $\bar{L}$. In the limit $\E \rightarrow 0$, decoherence acts to instantaneously select an eigenstate of the operator $\bar{L}$, and then freeze the quantum system in this eigenstate, via the quantum zeno effect \cite{misra1977zeno,gagen1993continuous,blanchard1993strongly}. In doing so, the quantum system is essentially classicalised, with any superpositions being supressed by the strong decoherence. Since the back-reaction on the classical system is determined by the eigenvalues of the operator $\bar{L}$, the classical system then undergoes deterministic evolution with drift given by the eigenstate that the quantum system is frozen in. Such a dynamics may be understood to be reversible on a subset of initial quantum states that are eigenstates of the Lindblad operator $\bar{L}$, but in general is highly non-deterministic, with a generic initial quantum state being rapidly decohered by the interaction.

The above example illustrates that in the $\E \rightarrow 0$ limit, dynamics arising from classical-quantum Hamiltonians that are self-commuting exhibit rapid decoherence in the quantum system. It turns out however that this is not a generic feature of dynamics in the $\E \rightarrow 0$ limit, which we may illustrate with the following example.

\textbf{A classical-quantum limit of two quantum harmonic oscillators.} Consider a system of two interacting quantum particles in one dimension, with the $Q$ subsystem characterised by the canonical commutation relation $[Q,P]=i\hbar$ and the $C$ subsystem as usual by $[\hat{q},\hat{p}]=i\hbar$. The system will be taken to have free evolution given by the bipartite quantum Hamiltonian $\hat{H}=\hat{p}^2/2m_C+ P^2/2m_Q+\lambda (\hat{q}-Q)^2$. Taking the classical-quantum limit of the $C$ subsystem gives a classical-quantum Hamiltonian $H=(p^2/2m_C)\mathds{1} + P^2/2m_Q + \lambda (q\mathds{1}-Q)^2$. For this model, one may compute the Lindblad and effective Hamiltonian operators of equations \eqref{eq: Lindblad_def} and \eqref{eq: Heff_def} exactly, exploiting the fact that the adjoint action closes under the set of linear operators in $\mathds{1},Q,P$ to obtain
\begin{equation} \label{eq: LqHO}
\begin{split}
L_{q}^H=\frac{ \E }{\hbar}\bigg[&  \sqrt{2\lambda m_Q}  \sin \left(\frac{\sqrt{2\lambda  } \hbar}{\E  \sqrt{m_Q}}\right) (q\mathds{1}-Q) \\
&+ [1-\cos \left(\frac{\sqrt{2\lambda } \hbar }{\E  \sqrt{m_Q}}\right)]P\bigg], 
\end{split}
\end{equation}
\begin{equation}
    L_{p}^H=\frac{p}{m_C}\mathds{1},
\end{equation}
\begin{equation} \label{eq: H_effHO}
\begin{split}
H_{eff}=&\frac{p}{2 m_C}\bigg[\frac{2\E m_Q}{\hbar} [\cos \left(\frac{\sqrt{2\lambda  } \hbar}{\E  \sqrt{m_Q}}\right) - 1]\\
&\quad +\sqrt{2\lambda m_Q} \sin \left(\frac{\sqrt{2\lambda  } \hbar}{\E  \sqrt{m_Q}}\right)\bigg] Q\\
&+\frac{p}{2 m_C}\bigg[1+ \cos\left(\frac{\sqrt{2\lambda  } \hbar}{\E  \sqrt{m_Q}}\right)  \\
&\quad-\frac{\sqrt{2m_Q}\E }{\sqrt{\lambda}\hbar} \sin \left(\frac{\sqrt{2\lambda  } \hbar}{\E  \sqrt{m_Q}}\right) \bigg] P,
\end{split}
\end{equation} where we have dropped terms proportional to the identity in $H_{eff}$. These explicit forms of Lindblad and effective Hamiltonian operators may be used in the master equation \eqref{eq: LcanCQ} or the unravelling equations \eqref{eq: unrav_C} and \eqref{eq: unrav_Q} to study the classical-quantum oscillator dynamics for arbitrary $\E $. 

Let us now consider this dynamics in the low diffusion, $\E \rightarrow0$ limit. The key feature of interest is that as $\E$ goes to zero, the non-trivial Lindblad operator $L_q^H$ responsible for quantum back-reaction on the classical system also vanishes. Moreover, one can check that the parts of $\bar{L}\varrho \bar{L}^\dag -\frac{1}{2}\{\bar{L}^\dag\bar{L},\varrho\}_+$ responsible for decoherence also vanish, at a faster rate than the rate at which the decoherence strength increases. In other words, in the $\E \rightarrow 0$ limit, both the back-reaction and decoherence of the quantum system are zero. To study the rest of the dynamics in the low diffusion limit, we first note that the remaining terms of  $\bar{L}\varrho \bar{L}^\dag -\frac{1}{2}\{\bar{L}^\dag\bar{L},\varrho\}_+$ lead to unitary evolution on the quantum system, with a Hamiltonian given by $(\hbar p/2 m_C \E)L_q^H$. Remarkably, although this Hamiltonian and $H_{eff}$ do not independently have well-defined $\E\rightarrow 0$ limits, their sum does and has its limit given by $(p/m_C)P$. We thus find that in the $E_f\rightarrow 0$ limit, the classical-quantum oscillator dynamics reduces to unitary dynamics on the quantum system under the classical-quantum Hamiltonian $H$ and an additional term $(p/m_C)P$, and the classical system experiences no back-reaction:
\begin{equation}
dq_t = \frac{p_t}{m_C} dt, \quad \quad \quad dp_t = 0,
\end{equation}
\begin{equation}
d|\psi\rangle_t = -\frac{i}{\hbar}(H + \frac{p_t}{m_C} P)|\psi\rangle_t dt.
\end{equation}
In this limit, the strong monitoring by an environment on the $C$ subsystem thus acts to effectively remove the back-reaction of the quantum system on the classical one, leaving simply coherent control of the quantum system by the classical one, despite the strength of interaction remaining fixed. Under the additional assumption that the $C$ subsystem is moving sufficiently slowly, one recovers coherent control under the classical-quantum Hamiltonian. This reproduces the results of earlier work on classical-quantum limits, which considered the special case in which the back-reaction is zero \cite{oliynyk2016classical,irish2022defining}. In our setting, this effect is reminiscent of dynamical decoupling, where the application of unitary pulses on a quantum system may reduce the interaction with an external environment \cite{viola1999dynamical}. Incidentally, one can check that the $\hbar \rightarrow 0$ limit of the operators defined in equations \eqref{eq: LqHO} and \eqref{eq: H_effHO} are still well-defined, and reduce the form of dynamics to that given in \eqref{eq: Lapprox}; the apparent difference in limiting behaviour as $\E $ becomes small is due to the non-commutativity of the two limits $\E \rightarrow 0$ and $\hbar \rightarrow 0$. 

The two examples above show that in the low diffusion limit, $\E \rightarrow 0$, the classical-quantum dynamics we find can exhibit two very different behaviours; one in which the quantum system rapidly decoheres, and affects the classical system, the other in which it evolves with unitary evolution, and has no back-reaction on the classical system. In the regime that a classical system appears to evolve without diffusion, it thus appears to be the case that any quantum degrees of freedom that are affecting the evolution of the system must be rapidly decohered and effectively classical, or do not influence the dynamics of the classical system, and undergo unitary evolution depending on the classical state of the system. It would be interesting to study how generic the latter case is, and indeed whether there exist other behaviours aside from these two.

\section{Discussion}

The main results, given in master equation form in \eqref{eq: LcanCQ} or stochastic unravelling form in \eqref{eq: unrav_C} and \eqref{eq: unrav_Q}, provide a physically motivated and consistent effective classical-quantum dynamics derived from a full quantum theory. A special limiting case of this, given in equation \eqref{eq: Lapprox}, provides a form of dynamics close to the quantum-classical Liouville equation that may be directly unravelled in classical trajectories in phase space and quantum trajectories in Hilbert space. Beyond the coupled quantum harmonic oscillators example given, understanding the kinds of dynamics obtained via this classical-quantum limit in further models, from optics to condensed matter theory, would be of great interest. With the form of Lindblad and effective Hamiltonian operators computed, the average and statistical properties of such systems may be numerically simulated using the stochastic unravellings of \eqref{eq: unrav_C} and \eqref{eq: unrav_Q}.

An important research direction to understand in greater detail is the conditions under which the above dynamics are a good approximation to a full quantum dynamics. While the work in this paper demonstrated that a classical-quantum limit gives a rich dynamical structure, the analysis was the classical-quantum analogue of the steps leading from the full Wigner dynamics of equation \eqref{eq: WigEvo} to the Liouville equation in \eqref{eq: C_liouville}. Understanding whether the various approaches that characterise the conditions under which one may make this approximation \cite{zurek1994decoherence,bhattacharya2000continuous,bhattacharya2003continuous,hernandez2023hbar,hernandez2023decoherence} can be generalised to the more complex case of a classical-quantum limit is an important open question.

The methods presented here rely on the assumption that the environment may be modelled in a particularly simple way, as a series of discrete-time decoherence channels on the subsystem that is classicalised. It would be interesting to understand whether the results we obtain here may also be derived directly from continuous-time models of an environment (c.f. Appendix \ref{app: cont}). Moreover, the effect of the environment in this proposed classical-limit procedure is particularly simplistic, characterised only by the total strength of phase space diffusion $\E$ and a parameter $s$ quantifying the relative strength of diffusion in position and momentum. In real systems, the environment may induce a large number of additional effects on the dynamics such as friction, and in such cases we expect the corresponding classical-quantum dynamics to be modified to reflect this. 

Beyond this, it would be interesting to understand dynamics which relax the requirement of preserving effective classical-quantum states. In its most modest form, this could arise from restricting the set of measurements performed on a system undergoing Markovian evolution, using the formalism of section \ref{sec: technic}, such that even in the presence of entanglement the system appears classical. More generally, dynamics that do not preserve effective classical-quantum states could arise analogously to how relaxing complete-positivity in the study of open quantum systems can sometimes approximate full non-Markovian dynamics more accurately than when the dynamics is of Lindblad form \cite{tanimura2015real}. In this regard, making precise the notions of {\it almost always classical-quantum}, or {\it approximately classical-quantum}, are likely to be important. 

For these reasons, the classical-quantum limit introduced here is likely to be one of many, and we anticipate that considerable further work is required to understand the full landscape and applications of effective classical-quantum dynamics beyond the theory we have presented.

\noindent \emph{Acknowledgements.}
We are indebted to Maite Arcos, Emanuele Panella, Juan Pedraza, Andrea Russo, Rob Spekkens, Andy Svesko and Zach Weller-Davies for many valuable discussions over the course of this work. Additionally, we would like to thank Andreas Dechant, Clemens Gneiting, Nicola Pranzini, Daniel Ranard, Keiji Saito, Toshihiro Yada and the attendees of the QIMG 2023 workshop YITP-T-23-01 for interesting conversations and feedback on an early draft of this work. JO is supported by an EPSRC Established Career Fellowship, and a Royal Society Wolfson Merit Award. I.L~acknowledges financial support from EPSRC. This research was supported by the National Science Foundation under Grant No. NSF PHY11-25915 and by the Simons Foundation {\it It from Qubit} Network.

\appendix
\numberwithin{equation}{section}

\section{Pawula and CQ Pawula theorems} \label{app: Pawula}
For convenience, we reproduce the two theorems relevant for characterising positivity of dynamics in classical limits, the Pawula theorem \cite{pawula1967generalizations} and the CQ Pawula theorem \cite{oppenheim2022two}, as well as explaining how the Liouville equation \eqref{eq: C_liouville}, quantum-classical Liouville equation \eqref{eq: CQ_liouville}, and classical-quantum generator \eqref{eq: LcanCQ} satisfy (or not) the various forms.

\textbf{Pawula (1957)} The general form of Markovian, linear, trace-preserving and positive dynamics of a real-valued function of phase space $P$ is either of Fokker-Planck form
\begin{equation} \label{eq: FP}
    \frac{\partial P}{\partial t}=  -\frac{\partial }{\partial z_i}(D_{1,i}  P ) +\frac{1}{2} \frac{\partial^2 }{\partial z_i \partial z_j} ( D_{2, i j}  P )\\
\end{equation} or it contains an infinite number of higher order derivative terms in phase space. The $i,j,\ldots$ indices run from $1$ to $n$, the number of phase space degrees of freedom $z_i$, and there is summation of repeated indices. Here, $D_{1,i}$ are the elements of a real vector of length $n$, $D_1$, and $D_{2,ij}$ are the elements of a real positive semi-definite $n\times n$ matrix $D_2$. All of the $D$ coefficients are allowed to have dependence on phase space.

\textbf{CQ Pawula (2022)} The general form of Markovian, linear, trace-preserving and completely-positive dynamics of an operator-valued function of phase space $\varrho$ is either of the form
\begin{equation} \label{eq: L_CP}
\begin{split}
\frac{\partial \varrho}{\partial t}=&  -\frac{\partial }{\partial z_i}(D_{1,i}^C  \varrho ) +\frac{1}{2} \frac{\partial^2 }{\partial z_i \partial z_j} ( D_{2, i j}  \varrho )\\
&-i[\bar{H},  \varrho ] + D_0^{\alpha \beta}\big( L_{\alpha}  \varrho L_{\beta}^{\dag} - \frac{1}{2}  \{ L_{\beta}^{\dag} L_{\alpha},  \varrho \}_+ \big) \\
&- \frac{\partial }{\partial z_{i}} \left( {D^{\alpha }_{1, i}}^* L_{\alpha}  \varrho +  \varrho D^{\alpha}_{1, i}  L_{\alpha}^{\dag} \right)  ,
\end{split}
\end{equation}
where 
\begin{align} \label{eq: pos}
    D_0 \succeq D_1^{\dag} D_2^{-1} D_1,\,\,\,\, (\mathbb{I}- D_2 D_2^{-1})D_1 =0, 
\end{align} or it contains an infinite number of higher order derivative terms in phase space. Here, the $i,j,\ldots$ indices run from $1$ to $n$, the number of phase space degrees of freedom $z_i$, while the $\alpha,\beta,\ldots$ indices run from $1$ to $p$, the number of traceless and orthogonal Lindblad operators $L_\alpha$ in Hilbert space. We assume summation over repeated indices of either kind. The various $D$ coefficients are organised as follows: $D_0^{\alpha \beta}$ are the elements of an $p\times p$ complex positive semi-definite matrix $D_0$, $D_{1,i}^{\alpha}$ are the elements of a complex $n\times p$ matrix $D_1$, which has conjugate transpose $D_1^\dag$, while ${D^{\alpha }_{1, i}}^*$ denotes the complex conjugate of $D_{1,i}^{\alpha}$. Additionally, $D_{1,i}^C$ are the elements of a real vector of length $n$, $D_1^C$, and $D_{2,ij}$ are the elements of a real positive semi-definite $n\times n$ matrix $D_2$, which has the generalised inverse $D_2^{-1}$. Finally, $\bar{H}$ is Hermitian operator. All the $D$ coefficients and $\bar{H}$ may have arbitrary dependence on $z$.

When the Lindblad operators are not chosen traceless and orthogonal, the above conditions on the dynamics can be shown to still be sufficient for complete-positivity, even when dependent on phase space.  In this case, the role of classical drift vector $D_1^C$ is essentially played by the component of the $L_\alpha$ proportional to the identity.

\subsection{Liouville equation}

The Liouville equation \eqref{eq: C_liouville} satisfies the Fokker-Planck form given by \eqref{eq: FP} for 
\begin{equation}
\begin{split}
D_{1,q}=\frac{\partial H}{\partial p}\quad& \quad D_{1,p}=-\frac{\partial H}{\partial q}\\
D_2&=\begin{pmatrix}
    0 & 0\\
    0 & 0    
    \end{pmatrix},
\end{split}
\end{equation} where here $H$ is the classical Hamiltonian.

\subsection{Quantum-classical Liouville equation}

The quantum-classical Liouville equation, when written in the form of \eqref{eq: L_CP} with phase space dependent Lindblad operators, has
\begin{equation}
\begin{split}
    L_q=\frac{\partial H}{\partial p},\quad L_p=-&\frac{\partial H}{\partial q},\quad \bar{H}=\frac{H}{\hbar}, \quad D_1^C=(0,0)^T \\ 
D_0=\begin{pmatrix}
    0 & 0\\
    0 & 0    
    \end{pmatrix} \quad 
D_1&=\begin{pmatrix}
    \frac{1}{2} & 0\\
    0 & \frac{1}{2}    
    \end{pmatrix} \quad 
D_2=\begin{pmatrix}
    0 & 0\\
    0 & 0    
    \end{pmatrix}.
\end{split}
\end{equation} where $H$ is the classical-quantum Hamiltonian. Since $D_2$ and $D_0$ are zero everywhere, but $D_1$ is not, the positivity conditions \eqref{eq: pos} are not satisfied, and thus the dynamics is not completely-positive.
 
\subsection{Classical-quantum dynamics of $\mathcal{L}$}

By the same reasoning as above, one may read from \eqref{eq: LcanCQ} the three matrices $D_0$, $D_1$ and $D_2$ given in \eqref{eq: D0},\eqref{eq: D1},\eqref{eq: D2} by taking the Lindblad operators to be $L_q^H$ and $L_p^H$. The remaining degrees of freedom are $\bar{H}=(H+H_{eff})/\hbar$ and $D_1^C=(0,0)^T$. Since $\bar{H}$ is Hermitian and the $D$ coefficients satisfy the requirements of \eqref{eq: pos}, the dynamics is completely-positive.

\section{Derivation of the generator $\mathcal{L}$} \label{app: L}

In this appendix, we provide the technical details needed to go from the dynamical map of equation \eqref{eq: totalW} to the form of generator given in equation \eqref{eq: Lseries} and equation \eqref{eq: LcanCQ}. 

To compute $\mathcal{L}$ given by equations \eqref{eq: Lseries} and \eqref{eq: LcanCQ} we first write out the evolution map $\mathcal{E}_\tau=\mathcal{E}_\tau^{\E \tau}$ explicitly as
\begin{widetext}
\begin{equation} \label{eq: dynamicMap}
    \mathcal{E}_\tau=e^{\frac{1}{2}(\frac{\E s^2}{2} \frac{\partial^2}{\partial q^2}  + \frac{\E }{2 s^2} \frac{\partial^2}{\partial p^2}) \tau} e^{-\frac{i}{\E }[H,\ \cdot \ ] + \frac{1}{2}(\{H,\ \cdot \ \} - \{\ \cdot \ ,H\})\tau + O(\tau^2)} e^{\frac{1}{2}(\frac{\E s^2}{2} \frac{\partial^2}{\partial q^2}  + \frac{\E }{2 s^2} \frac{\partial^2}{\partial p^2}) \tau}.
\end{equation}
The most important part of this expression to notice immediately is that the first term in the middle exponential has no $\tau$ dependence – this term is ultimately responsible for most of the subsequent structure of the generator $\mathcal{L}$.

To go from this dynamical map to a generator, we will use equation \eqref{eq: Ldef}. This expression generalises the standard method for constructing generators to the case where the dynamical maps which do not reduce to the identity map at $\tau = 0$. This arises by identifying two separate contributions to the short-time dynamics, $\mathcal{L}_1$ and $\mathcal{L}_2$, corresponding to the first and second terms of \eqref{eq: Ldef}. The first part $\mathcal{L}_1$ can be understood as the standard one, defined as $\mathcal{L}_1=\lim_{\tau\rightarrow 0} \mathcal{L}_{1,\tau}$, where $\mathcal{L}_{1,\tau}\varrho(\tau)=\partial_\tau \varrho(\tau)$. This describes how the state of the system changes from $t=0$ to a short time $t=\tau$ later as $\tau \rightarrow 0$. This may be computed from the dynamical map by noting that $\varrho(\tau)=\mathcal{E}_\tau \varrho(0)$ and $\varrho(0)=\mathcal{E}_\tau^{-1} \varrho(\tau)$, which may be substituted into the definition of $\mathcal{L}_{1,\tau}$ to arrive at the first term of \eqref{eq: Ldef}. The second part of the generator $\mathcal{L}_2$ can be understood as arising from the part of the dynamical map at $\tau=0$, which in contrast to standard treatments, is not proportional to the identity. That the map is not proportional to the identity here occurs as a result of inadvertently taking $\hbar \rightarrow 0$ even for purely quantum degrees of freedom, and so should be corrected for by restoring $E_f=\hbar/\tau$.

We first turn to explicitly finding an expression for the second term of equation \eqref{eq: Ldef}, and providing further justification for its appearance, using equation \eqref{eq: dynamicMap}. As described above, this term arises from the part of the dynamical map at $\tau=0$, which is given
\begin{equation}
 \mathcal{E}_0 =e^{-\frac{i}{\E } [H,\ \cdot\ ]}.
\end{equation} After $N=t/\tau$ evolution steps, the total contribution of this part of the dynamics is
\begin{equation}
\mathcal{E}_0^N=\mathcal{E}_0^{t/\tau}=e^{-\frac{i}{\E  \tau} [H,\ \cdot\ ] t}
\end{equation} and thus is generated by the unitary term $-\frac{i}{\hbar}[H,\ \cdot\ ]$ if we restore $\hbar=\E \tau$. This agrees with the expression given by the second term of equation \eqref{eq: Ldef}. Although in principle the unitary steps occur in between steps generated by the $\tau$-dependent part of the generator, any changes to the generator from the non-commutativity of these terms are of $O(\tau)$, and thus vanish in the $\tau\rightarrow 0$ limit, meaning that the resulting dynamics is captured by the generator $\mathcal{L}_2=-\frac{i}{\hbar}[H,\ \cdot\ ]$.

To compute the first part of the generator $\mathcal{L}$, denoted above as $\mathcal{L}_1$, we take the derivative of $\mathcal{E}_\tau$ to give 
\begin{equation}
\begin{split}
    \frac{\partial }{\partial \tau } \mathcal{E}_\tau =&\frac{1}{2}\big(\frac{\E s^2}{2} \frac{\partial^2}{\partial q^2} + \frac{\E }{2 s^2} \frac{\partial^2}{\partial p^2} \big) \mathcal{E}_\tau\\
    &+ e^{\frac{\E }{2} D \tau}\frac{e^{\ad{ \frac{-i}{\E } [H,\cdot]+O(\tau)}}-1}{\ad{ \frac{-i}{\E } [H,\cdot]+O(\tau)}}\bigg( \frac{1}{2}\{H,\cdot\} - \frac{1}{2}\{\cdot,H\}+ O(\tau) \bigg)e^{-\frac{\E }{2} D \tau}\mathcal{E}_\tau \\
    &+ \frac{1}{2}\mathcal{E}_\tau \big(\frac{\E s^2}{2} \frac{\partial^2}{\partial q^2} + \frac{\E }{2 s^2} \frac{\partial^2}{\partial p^2} \big),
\end{split}
\end{equation}
where here $D=\frac{\E s^2}{2} \frac{\partial^2}{\partial q^2} + \frac{\E }{2 s^2} \frac{\partial^2}{\partial p^2} $. This gives the first component of the generator $\mathcal{L}$ as
\begin{equation}
\begin{split}
    \lim_{\tau\rightarrow 0} \big(\frac{\partial }{\partial \tau } \mathcal{E}_\tau \big)\mathcal{E}_\tau^{-1} =&\frac{1}{2}\big(\frac{\E s^2}{2} \frac{\partial^2}{\partial q^2} + \frac{\E }{2 s^2} \frac{\partial^2}{\partial p^2} \big) \\
    &+ \frac{e^{\ad{ \frac{-i}{\E } [H,\cdot]}}-1}{\ad{ \frac{-i}{\E } [H,\cdot]}}\bigg( \frac{1}{2}\{H,\cdot\} - \frac{1}{2}\{\cdot,H\} \bigg) \\
    &+ \frac{1}{2}e^{-\frac{i}{\E }[H,\ \cdot \ ]} \big(\frac{\E s^2}{2} \frac{\partial^2}{\partial q^2} + \frac{\E }{2 s^2} \frac{\partial^2}{\partial p^2} \big) e^{\frac{i}{\E }[H,\ \cdot \ ]}
\end{split}
\end{equation}
where the $O(\tau)$ terms disappear in the $\tau\rightarrow 0$ limit and we have used the fact that $\lim_{\tau\rightarrow 0}e^{\pm\frac{\E }{2} D \tau}$ is the identity operator. Noting that we may use the following equality between the exponential of the adjoint and the adjoint of the exponential
\begin{equation} \label{eq: adAd}
    e^{\ad{\mathcal{B}}}\mathcal{A}=e^{\mathcal{B}}\mathcal{A}e^{-\mathcal{B}}
\end{equation} we find the quoted form of the generator in equation \eqref{eq: Lseries}.

To compute the form of the generator given in \eqref{eq: LcanCQ} is a little more work.  Denoting the following term $\mathcal{T}_1$
\begin{equation}
    \mathcal{T}_1\varrho=\frac{1}{2}e^{\ad{ \frac{-i}{\E } [H,\cdot]}} \bigg( \frac{\E s^2}{2}\frac{\partial^2  }{\partial q^2}+ \frac{\E }{2s^2}\frac{\partial^2 }{\partial p^2} \bigg) \varrho
\end{equation} where we introduce an aribtrary CQ state $\varrho$ to make the action of this generator explicit, one may use the equality between the exponential of the adjoint and the adjoint of the exponential as in equation \eqref{eq: adAd} to rewrite this as
\begin{equation}
    \frac{1}{2}e^{-\frac{i}{\E }[H,\ \cdot \ ]} \big(\frac{\E s^2}{2} \frac{\partial^2}{\partial q^2} + \frac{\E }{2 s^2} \frac{\partial^2}{\partial p^2} \big) e^{+\frac{i}{\E }[H,\ \cdot \ ]}\varrho
\end{equation} and then use it \textit{again}, noting that $\pm\frac{i}{\E }[H,\dt]=\ad{\frac{\pm i}{\E }H}$, to give
\begin{equation}
    \frac{1}{2}e^{\frac{-i}{\E }H} \big(\frac{\E s^2}{2} \frac{\partial^2}{\partial q^2} + \frac{\E }{2 s^2} \frac{\partial^2}{\partial p^2} \big) (e^{\frac{i}{\E }H}\varrho e^{\frac{-i}{\E }H})e^{\frac{i}{\E }H}.
\end{equation} One may then compute this expression explicitly, taking care to note that whenever a derivative is made of the exponential of a $z=q,p$ dependent operator, that
\begin{equation}
\begin{split}
    \frac{\partial}{\partial z}e^{-\frac{i}{\E }H}&=-\frac{i}{\E }\frac{e^{\ad{ \frac{-i}{\E } H}}-1}{\ad{ \frac{-i}{\E } H}}  \bigg(\frac{\partial H}{\partial z}\bigg) e^{-\frac{i}{\E }H} \\
    &= -\frac{i}{\E } L^H_z e^{-\frac{i}{\E }H},
\end{split}
\end{equation} and
\begin{equation}
\begin{split}
    \frac{\partial}{\partial z}e^{\frac{i}{\E }H}&=\frac{i}{\E } e^{\frac{i}{\E }H}\  \frac{e^{\ad{ \frac{-i}{\E } H}}-1}{\ad{ \frac{-i}{\E } H}}  \bigg(\frac{\partial H}{\partial z}\bigg) \\
    &= \frac{i}{\E }  e^{\frac{i}{\E }H} L^H_z.
\end{split}
\end{equation} Using these formulae, one may show that
\begin{equation}
\begin{split}
    \frac{1}{2}e^{-\frac{i}{\E }H} \frac{\partial^2}{\partial z^2} (e^{\frac{i}{\E }H}\varrho e^{-\frac{i}{\E }H})e^{\frac{i}{\E }H}
    =&- \frac{i}{2\E }[\frac{\partial L_z^H}{\partial z},\varrho]+\frac{i}{\E }\frac{\partial}{\partial z}[L_z^H,\varrho]\\
    &+\frac{1}{\E ^2}\big(L_z^H \varrho L_z^H - \frac{1}{2}\{{L_z^H}^2,\varrho\}_+\big) + \frac{1}{2}\frac{\partial^2 \varrho}{\partial z^2} ,
\end{split}
\end{equation} which gives the overall generator $\mathcal{T}_1$ as
\begin{equation}
    \begin{split}
        \mathcal{T}_1\varrho=&-i[\frac{s^2}{4}\frac{\partial L_q^H}{\partial q} + \frac{1}{4s^2}\frac{\partial L_p^H}{\partial p},\varrho]+\frac{i s^2}{2}\frac{\partial}{\partial q}[L_q^H,\varrho]+\frac{i}{2s^2}\frac{\partial}{\partial p}[L_p^H,\varrho] \\
        &+\frac{s^2}{2\E }(L_q^H \varrho L_q^H - \frac{1}{2}\{{L_q^H}^2,\varrho\}_+) + \frac{1}{2\E s^2}(L_p^H \varrho L_p^H - \frac{1}{2}\{{L_p^H}^2,\varrho\}_+)\\
        &+ \frac{1}{2}\big(\frac{\E s^2}{2}\frac{\partial^2 \varrho}{\partial q^2}+ \frac{\E }{2s^2}\frac{\partial^2 \varrho}{\partial p^2}\big) .
    \end{split}
\end{equation} 

The other component of $\mathcal{L}$ that remains to be computed we will denote $\mathcal{T}_2$ and is given
\begin{equation} \label{eq: T2}
    \begin{split}
        \mathcal{T}_2 \varrho =\frac{e^{\ad{ \frac{-i}{\E } [H,\cdot]}}-1}{\ad{ \frac{-i}{\E } [H,\cdot]}}\bigg( \frac{1}{2}\{H,\dt\} - \frac{1}{2}\{\dt,H\} \bigg)\varrho .
    \end{split}
\end{equation} Since the fraction $\frac{e^\ad{}-1}{\ad{}}$ is to be interpreted as describing a power series, and using the symmetry of second derivatives of $H$ to rewrite the Alexandrov-bracket as the derivatives of anticommutators, we may rewrite this generator more explicitly as
\begin{equation} \label{eq: T2_series}
    \begin{split}
        \mathcal{T}_2 \varrho=\sum_{n=0}^\infty \frac{1}{(n+1)!}
        \ad{\frac{-i}{\E } [H,\cdot]}^n \bigg( -\frac{1}{2}\frac{\partial}{\partial q}\{\frac{\partial H}{\partial p}, \dt\}_+ 
        +\frac{1}{2}\frac{\partial}{\partial p}\{\frac{\partial H}{\partial q} ,\dt\}_+ \bigg)\varrho .
    \end{split}
\end{equation}
To compute this infinite series, we will first need to find the commutation relations of the algebra generated by $-\frac{i}{\E } [H,\cdot]$, as one would do for the case for a Lie algebra of a Lie group – for some related work in the purely quantum case, see \cite{machnes2014surprising}. To simplify this subsequent analysis, we will use a shorthand $\mathbb{L}(A,\dt,B)$ to denote a generic component of a Lindblad decoherence generator i.e.
\begin{equation}
    \mathbb{L}(A,\varrho,B)=A\varrho B - \frac{1}{2} B A \varrho - \frac{1}{2} \varrho B A.
\end{equation} One may then compute the commutation relations between $-\frac{i}{\E } [H,\cdot]$ and the various terms that appear: (a) with the derivative of an anticommuator 
\begin{equation} \label{eq: adH_derivative_anticom}
    \ad{\frac{-i}{\E } [H,\dt]} \ \frac{\partial}{\partial z}\{A ,\dt\}_+ = \frac{\partial}{\partial z}\{\ad{\frac{-i}{\E }H}A ,\dt\}_+ + \frac{i}{\E } \mathbb{L}(\frac{\partial H}{\partial z},\dt,A) - \frac{i}{\E } \mathbb{L}(A,\dt,\frac{\partial H}{\partial z}) + \frac{i}{\E }[\frac{1}{2}\{ A,\frac{\partial H}{\partial z}\}_+,\dt],
\end{equation} 
(b) with the component of the Lindblad decoherence generator
\begin{equation}\label{eq: adH_Lindblad}
\begin{split}
    \ad{\frac{-i}{\E } [H,\dt]}\mathbb{L}(A,\dt,B)=&\mathbb{L}(\ad{\frac{-i}{\E }H}A,\dt,B)
    +\mathbb{L}(A,\dt,\ad{\frac{-i}{\E }H}B),
\end{split}
\end{equation}
and (c) with a unitary generator
\begin{equation}\label{eq: adH_unitary}
    \ad{\frac{-i}{\E } [H,\dt]} -i[A,\dt]=-i[\ad{\frac{-i}{\E }H}A,\dt].
\end{equation} Since the above generators are closed under the repeated action of $\ad{\frac{-i}{\E } [H,\dt]} $, these relations are sufficient to compute the above series. 

To actually compute the series, we will consider each kind of generator (a)-(c) separately. Starting with (a), the derivative of an anticommutator, we note that the adjoint action of the commutator with $H$ is equivalent to the adjoint action with $H$ on the operator in question. This gives the first part of $\mathcal{T}_2$ as
\begin{equation} \label{eq: T2_a}
\begin{split}
\mathcal{T}_2^{(a)}&=\sum_{n=0}^\infty \frac{1}{(n+1)!} \bigg(-\frac{1}{2}\frac{\partial}{\partial q}\{\ad{\frac{-i}{\E }H}^n\frac{\partial H}{\partial p}, \dt\}_+ +\frac{1}{2}\frac{\partial}{\partial p}\{\ad{\frac{-i}{\E }H}^n\frac{\partial H}{\partial q} ,\dt\}_+\bigg)\\
    &=-\frac{1}{2}\frac{\partial}{\partial q}\{L_p^H, \dt\}_+ +\frac{1}{2}\frac{\partial}{\partial p}\{L_q^H ,\dt\}_+ ,
\end{split}
\end{equation} where we have again used the series expansion of $\frac{e^\ad{}-1}{\ad{}}$. 

To compute $\mathcal{T}_2^{(b)}$, the part corresponding to the Lindblad terms, we first write down the form of the $O(\E ^{-n})$ order term, which is given
\begin{equation} \label{eq: order_n_Lindblad}
    \frac{i}{2\E }\sum_{k=0}^{n-1}\mathbb{L}(\frac{1}{(k+1)!}\ad{\frac{-i}{\E }H}^k \frac{\partial H}{\partial p},\dt,\frac{1}{(n-k)!}\ad{\frac{-i}{\E }H}^{n-1-k} \frac{\partial H}{\partial q}) - \mathbb{L}(\frac{1}{(k+1)!}\ad{\frac{-i}{\E }H}^k \frac{\partial H}{\partial q},\dt,\frac{1}{(n-k)!}\ad{\frac{-i}{\E }H}^{n-1-k} \frac{\partial H}{\partial p})
\end{equation} 
 We will now show by induction that this is true for all $n\geq1$. When $n=1$, all the Lindblad terms come from the application of \eqref{eq: adH_derivative_anticom} on the Alexandrov-bracket, which one can check agrees with the above expression (being careful to include the factor of $1/(1+1)!$ coming from \eqref{eq: T2_series}). For an arbitrary term of order $n+1$, it follows from \eqref{eq: adH_derivative_anticom}-\eqref{eq: adH_unitary} that all terms must come from the application of $\frac{1}{n+2}\ad{\frac{-i}{\E } [H,\dt]} $ to either the previous $n$th order term in \eqref{eq: order_n_Lindblad} or to the $n$th order term of \eqref{eq: T2_a}. This gives the $(n+1)$th order term in total as
\begin{equation}
\begin{split}
\frac{i}{2\E }\frac{1}{n+2}\sum_{k=0}^{n-1}\bigg\{
&\mathbb{L}(\frac{1}{(k+1)!}\ad{\frac{-i}{\E }H}^{k+1} \frac{\partial H}{\partial p},\dt,\frac{1}{(n-k)!}\ad{\frac{-i}{\E }H}^{n-1-k} \frac{\partial H}{\partial q}) \\
+&\mathbb{L}(\frac{1}{(k+1)!}\ad{\frac{-i}{\E }H}^k \frac{\partial H}{\partial p},\dt,\frac{1}{(n-k)!}\ad{\frac{-i}{\E }H}^{n-k} \frac{\partial H}{\partial q}) \\
-& \mathbb{L}(\frac{1}{(k+1)!}\ad{\frac{-i}{\E }H}^{k+1} \frac{\partial H}{\partial q},\dt,\frac{1}{(n-k)!}\ad{\frac{-i}{\E }H}^{n-1-k} \frac{\partial H}{\partial p})\\
-& \mathbb{L}(\frac{1}{(k+1)!}\ad{\frac{-i}{\E }H}^k \frac{\partial H}{\partial q},\dt,\frac{1}{(n-k)!}\ad{\frac{-i}{\E }H}^{n-k} \frac{\partial H}{\partial p}) \bigg\} \\
+\frac{i}{2\E }\frac{1}{(n+2)!}\bigg\{
&\mathbb{L}(\frac{\partial H}{\partial p},\dt,\ad{\frac{-i}{\E }H}^{n} \frac{\partial H}{\partial q})+\mathbb{L}(\ad{\frac{-i}{\E }H}^{n}\frac{\partial H}{\partial p},\dt, \frac{\partial H}{\partial q}) \\
-&\mathbb{L}(\frac{\partial H}{\partial q},\dt,\ad{\frac{-i}{\E }H}^{n} \frac{\partial H}{\partial p})-\mathbb{L}(\ad{\frac{-i}{\E }H}^{n}\frac{\partial H}{\partial q},\dt, \frac{\partial H}{\partial p}) \bigg\}.
\end{split}
\end{equation} Considering first the Lindblad terms with one entry $\ad{}^n$ and the other $\ad{}^0$, we see that the numerical prefactors of these terms are given
\begin{equation}
    \frac{1}{n+2}\frac{1}{n!}+ \frac{1}{(n+2)!}=\frac{1}{(n+1)!},
\end{equation} with the term on the left hand side coming from $k=n-1$ or $0$ terms, and the right hand side coming from the bottom two lines. Analogously, for a generic Lindblad term with one entry $\ad{}^m$ and the other $\ad{}^{n-m}$ for $0< m < n$ we have two terms coming from the sum over $k$, given
\begin{equation}
\frac{1}{n+2}\bigg(\frac{1}{m!}\frac{1}{(n-m+1)!}+\frac{1}{(m+1)!}\frac{1}{(n-m)!}\bigg)=\frac{1}{(m+1)!(n+1-m)!},
\end{equation} which implies that the $(n+1)$th order terms may be written as
\begin{equation}
    \frac{i}{2\E }\sum_{k=0}^{n}\mathbb{L}(\frac{1}{(k+1)!}\ad{\frac{-i}{\E }H}^k \frac{\partial H}{\partial p},\dt,\frac{1}{(n+1-k)!}\ad{\frac{-i}{\E }H}^{n-k} \frac{\partial H}{\partial q}) - \mathbb{L}(\frac{1}{(k+1)!}\ad{\frac{-i}{\E }H}^k \frac{\partial H}{\partial q},\dt,\frac{1}{(n+1-k)!}\ad{\frac{-i}{\E }H}^{n-k} \frac{\partial H}{\partial p})
\end{equation} which indeed is the expression \eqref{eq: order_n_Lindblad} with $n\rightarrow n+1$. Since this expression is only the $n$th order term, we may write $\mathcal{T}_2^{(b)}$ as the sum over all these terms
\begin{equation} \label{eq: T2_b}
    \frac{i}{2\E }\sum_{n=1}^\infty\sum_{k=0}^{n-1}\mathbb{L}(\frac{1}{(k+1)!}\ad{\frac{-i}{\E }H}^k \frac{\partial H}{\partial p},\dt,\frac{1}{(n-k)!}\ad{\frac{-i}{\E }H}^{n-1-k} \frac{\partial H}{\partial q}) - \mathbb{L}(\frac{1}{(k+1)!}\ad{\frac{-i}{\E }H}^k \frac{\partial H}{\partial q},\dt,\frac{1}{(n-k)!}\ad{\frac{-i}{\E }H}^{n-1-k} \frac{\partial H}{\partial p})
\end{equation}  which noting that $\mathbb{L}$ is linear each of its arguments can be simplified to 
\begin{equation}
    \mathcal{T}_2^{(b)}= \frac{i}{2\E }\mathbb{L}(L_p^H,\dt,L_q^H)-\frac{i}{2\E }\mathbb{L}(L_q^H,\dt,L_p^H)
\end{equation} i.e.
\begin{equation}
\begin{split}
    \mathcal{T}_2^{(b)}\varrho= -\frac{i}{2\E }(L_q^H \varrho L_p^H - \frac{1}{2}\{L_p^H L_q^H,\varrho\}_+) + \frac{i}{2\E }(L_p^H \varrho L_q^H - \frac{1}{2}\{L_q^H L_p^H,\varrho\}_+).
\end{split}
\end{equation} \end{widetext}

The final component of $\mathcal{T}_2$ to compute is the unitary part, which we will keep track of by defining an associated Hamiltonian $H_{qp+pq}$ via $\mathcal{T}_2^{(c)}=-i[H_{qp+pq},\dt]$. From \eqref{eq: adH_derivative_anticom}-\eqref{eq: adH_unitary} it is apparent that any contributions to $H_{qp+pq}$ are generated by the action of $\ad{\frac{-i}{\E } [H,\dt]}$ on derivatives of anticommutator terms, given by \eqref{eq: T2_a}, and then the subsequent action of $\ad{\frac{-i}{\E } [H,\dt]}$ on the unitary terms generated by these. The numerical factor coming from the repeated action in \eqref{eq: T2_series} may be kept track of by simply noting that the $O(\E ^{-n})$ terms have a factor $1/(n+1)!$. This lets us write down the Hamiltonian $H_{qp+pq}$ as
\begin{equation}
\begin{split}
    H_{qp+pq}=\frac{1}{4\E }&\sum_{n,m=0}^\infty \frac{1}{(n+m+2)!} \\
    &\times \ad{\frac{-i}{\E }H}^{n}\bigg[\{\ad{\frac{-i}{\E }H}^{m}\frac{\partial H}{\partial p},\frac{\partial H}{\partial q}\}_+\\
    &-\{\ad{\frac{-i}{\E }H}^{m}\frac{\partial H}{\partial q},\frac{\partial H}{\partial p}\}_+\bigg],
    \end{split}
\end{equation} with the sum over $m$ indicating the initial creation of a unitary term via \eqref{eq: adH_derivative_anticom}, and the sum over $n$ giving the subsequent action via \eqref{eq: adH_unitary}. Using the fact that 
\begin{equation}
    \ad{\frac{-i}{\E }H}^n\{A,B\}_+=\sum_{k=0}^n \binom{n}{k} \{\ad{\frac{-i}{\E }H}^{n-k}A,\ad{\frac{-i}{\E }H}^k B\}_+
\end{equation} where $\binom{n}{k}$ is the binomial coefficient, and collecting terms, we finally arrive at the form 
\begin{equation}
\begin{split}
    H_{qp+pq}=\frac{1}{4\E }&\sum_{n,m=0}^\infty \frac{C_{nm}}{(n+m+2)!}\ \\
    &\times \{ \ad{\frac{-i}{\E }H}^n \frac{\partial H}{\partial q},\ad{\frac{-i}{\E }H}^m \frac{\partial H}{\partial p} \}_+ .
\end{split}
\end{equation}
Here the coefficients $C_{nm}$, explicitly given by
\begin{equation}
    C_{nm}= \sum_{r=0}^m\frac{(r+n)!}{r!n!} - \sum_{r=0}^n\frac{(r+m)!}{r!m!},
\end{equation} may be written out pictorially to show that they are generated by a version of the Pascal triangle, here with the same addition rules but with the boundary elements given by the integers $\mathbb{Z}$ i.e.

\begin{tabular}{>{}l<{\hspace{12pt}}*{13}{c}}
&&&&&&&0&&&&&&\\
&&&&&&-1&&\ 1&&&&&\\
&&&&&-2&&0&&\ 2&&&&\\
&&&&-3&&-2&&\ 2&&\ 3&&&\\
&&&-4&&-5&&0&&\ 5&&\ 4&&\\
&&-5&&-9&&-5&&\ 5&&\ 9&&\ 5&\\
&-6&&-14&&-14&&0&&\ 14&&\ 14&&\ 6\\
\\
\end{tabular}

Finally, combining the components $-\frac{i}{\hbar}[H, \dt]$, $\mathcal{T}_1$ and $\mathcal{T}_2$, using the definition
\begin{equation}
   H_{eff}=\frac{\hbar s^2}{4}\frac{\partial L_q^H}{\partial q}+\frac{\hbar}{4s^2}\frac{\partial L_p^H}{\partial p} + \hbar H_{qp+pq}
\end{equation} gives the form quoted in \eqref{eq: LcanCQ}.

\section{Including $O(\hbar)$ contributions in the classical-quantum Hamiltonian} \label{app: O(hbar)}

If instead of assuming $H^W=H+O(\hbar^2)$ we had assumed $H^W=H+\hbar H^1+ O(\hbar^2)$, we would find that the equations of motion for the Liouville equation are unchanged, but that there is a change in the quantum-classical Liouville equation. Specifically, the $O(\hbar^0)$ part of the partial Wigner generator now takes the form
\begin{equation}
\begin{split}
    \mathcal{L}^W  =& -\frac{i}{\hbar}[H,\dt] - i[H^1,\dt] \\
    &+\frac{1}{2}(\{H, \dt\} - \{\dt,H\}) + O(\hbar).
\end{split}
\end{equation} Following the same steps as before in computing the generator, the only change is found at the level of the $\mathcal{T}_2$ component given in \eqref{eq: T2}, which now has the additional term $\mathcal{T}_2^1$
\begin{equation}
\mathcal{T}_2^1=\frac{e^{\ad{ \frac{-i}{\E } [H,\cdot]}}-1}{\ad{ \frac{-i}{\E } [H,\cdot]}}\bigg( -i[H^1,\dt] \bigg).
\end{equation} To put this in canonical form, we note that $\ad{[A,\dt]} [B,\dt]= [\ad{A} B,\dt]$, which follows from the Jacobi identity, and thus using the series expansion of $\frac{e^x-1}{x}$ and resumming we find
\begin{equation}
\mathcal{T}_2^1=-i[\frac{e^{-\frac{i}{\E } H}-1}{\ad{ \frac{-i}{\E } H}}H^1,\dt] .
\end{equation} Considering an $H^W$ with $O(\hbar)$ terms thus simply leads to an additional unitary term, and does not affect the resulting complete-positivity of the dynamics.

\section{Double scaling limit of a continuous Lindbladian evolution} \label{app: cont}

In this appendix we consider a double scaling limit of a continuous time Lindbladian dynamics, and illustrate the difference with the classical-quantum limit we present. In particular, we show that while this set-up can reproduce the stochastic classical dynamics of \eqref{eq: L_C}, it does not coincide with \eqref{eq: LcanCQ} and in fact does not describe completely-positive dynamics. This example illustrates that one must be careful in constructing classical-quantum limits simply by identifying scaling limits that leave diffusion in the classical degrees of freedom.

To start with, consider the following model of bipartite dynamics on the $C$ and $Q$ subsystems
\begin{equation}
\begin{split}
    \frac{\partial \hat{\rho}}{\partial t}=&-\frac{i}{\hbar}[\hat{H},\hat{\rho}] \\
    &+ \frac{\gamma}{s^2} (\hat{q} \hat{\rho} \hat{q}-\frac{1}{2}\{\hat{q}^2,\hat{\rho}\}_+)\\
    &+ \gamma s^2 (\hat{p} \hat{\rho} \hat{p}-\frac{1}{2}\{\hat{p}^2,\hat{\rho}\}_+).
\end{split}
\end{equation} where $\gamma$ is positive parameter controlling the overall rate of decoherence, while $s$ is a positive parameter controlling the relative strength of decoherence between position and momentum. This model is of the GKSL form \cite{lindblad1976generators,gorini1976completely}, and is thus completely-positive at the level of the quantum dynamics. To consider how this dynamics appears in the partial Wigner representation, we first rewrite each Lindblad term as a double commutator with $-i/\hbar$ prefactors i.e.
\begin{equation}
    \hat{z}\hat{\rho}\hat{z}-\frac{1}{2}\{\hat{z}^2,\hat{\rho}\}_+=\frac{\hbar^2}{2}\big(-\frac{i}{\hbar}[\hat{z},\ -\frac{i}{\hbar}[\hat{z},\hat{\rho}]\ ]\big),
\end{equation} for $\hat{z}=\hat{q},\hat{p}$. Using the mapping of operators to the partial Wigner representation $\hat{f} \hat{g}\mapsto f\star g$, which here amounts to identifying the commutators with $-i/\hbar$ prefactors with Poisson brackets, one arrives at the partial Wigner representation of this dynamics
\begin{equation}
\begin{split}
    \frac{\partial \varrho^W}{\partial t}=&-\frac{i}{\hbar} [H,\varrho^W]+\frac{1}{2}(\{H,\varrho^W\} - \{\varrho^W,H\})\\
    &+ \frac{\gamma \hbar^2}{2s^2}\{q,\{q,\varrho^W\}+ \frac{\gamma \hbar^2 s^2}{2}\{p,\{p,\varrho^W\} \\&+ O(\hbar),
\end{split}
\end{equation} where the $O(\hbar)$ terms are those truncated in \eqref{eq: masterL_W} to arrive at \eqref{eq: CQ_liouville}. 

In this case, the natural limit to take if one wishes to remove the $O(\hbar)$ terms and additionally preserve the terms containing $\gamma$ is the double scaling limit 
\begin{equation}
    \hbar\rightarrow 0,\ \gamma\rightarrow \infty\ \ \  s.t. \ \ \ \gamma \hbar^2 = \E.
\end{equation} Doing so with the above dynamics whilst naively ignoring $O(\hbar^{-1})$ terms gives the final form of master equation as
\begin{equation}
\begin{split}
    \frac{\partial \varrho^W}{\partial t}=&-\frac{i}{\hbar}[H,\varrho^W]+\frac{1}{2}(\{H,\varrho^W\} - \{\varrho^W,H\}) \\
    &+ \frac{\E}{2s^2}\frac{\partial^2 \varrho^W}{\partial p^2} + \frac{\E s^2}{2}\frac{\partial^2 \varrho^W}{\partial q^2}.
\end{split}
\end{equation} 

At first glance, this appears to be a valid classical-quantum limit incorporating the affects of decoherence. Indeed, substituting in a trivial Hamiltonian $H=H(q,p) \mathds{1}$ as in \ref{sec: two}, one finds that this dynamics reduces exactly to the stochastic Liouville dynamics given in \eqref{eq: L_C}. However, comparing the form of this dynamics to the general form of completely-positive Markovian classical-quantum dynamics given in Appendix \ref{app: Pawula}, one sees that the dynamics is in fact not a valid dynamics; it has non-zero back-reaction and diffusion, but without any decoherence on the quantum degrees of freedom. Moreover, repeating the above steps in either the partial Husimi $\varrho^Q$ or the partial Glauber-Sudarshan $\varrho^P$ representations using \eqref{eq: L_Q} and \eqref{eq: L_P}, one can see that the resulting form of dynamics is dependent on the choice of phase space representation used, even in the $\hbar \rightarrow 0$ limit. This example suggests some importance of the specific model of decoherence that is used to construct the classical-quantum limit we propose.

\section{Effective classicality of states and measurements} \label{app: effClass}

In this appendix, we summarise the framework of characterising the nonclassicality of a quantum system in terms of its states and measurements \cite{spekkens2008negativity,bartlett2012reconstruction}. The main discussion of \ref{sec: technic} may then be understood as the generalisation of this to the case of a classical \textit{sub}system. 

Considering here a single quantum system denoted $C$ as in section \ref{sec: wig}, we note that the general framework for discussing measurements is that of POVMs $\{\hat{E}_i\}$, where $\hat{E}_i$ denote the POVM elements \cite{nielsen2002quantum}. A quasiprobability representation $R$ is the assignment to every state $\hat{\rho}$ and every set of POVM elements $\hat{E}_i$ the real-valued functions of phase space $R(z)$ and $E^R_i(z)$ respectively, such that
\begin{equation} \label{eq: Rep}
    \tr[\hat{\rho}\hat{E_i}]=\int dz\  R (z) E^R_i(z).
\end{equation} This allows for the statistics of every measurement performed on the quantum system to be represented in terms of functions in phase space, with the Wigner representation introduced in section \ref{sec: wig} providing one such example. As in the case of the Wigner representation, the functions $R(z)$ and $E^R_i(z)$ will not be positive for all states and measurements, and thus cannot always be interpreted as a classical probability distribution with an associated distribution over classical measurement outcomes. However, when a restricted set of states $\{\hat{\rho}_\lambda\}$ and measurements $\{\{\hat{E}_i\},\{\hat{F}_i\},\ldots\}$ are considered, it turns out it is possible to find subtheories for which both states and measurements are represented positively. In particular, we will say that \textit{a set of states and POVMs admit a classical description} when there exists a representation $R$ such that $R_\lambda(z)$ and $E^R_i(z),F^R_i(z),\ldots$ are non-negative for all $z$. Important examples of subtheories that admit classical descriptions are the measurements and states associated with Gaussian quantum optics \cite{bartlett2012reconstruction,serafini2017quantum} and stabilizer circuits in quantum computing \cite{gottesman1998heisenberg,aaronson2004improved}.

The above framework allows one to distinguish when a set of states and measurements in a quantum theory can be explained by a classical one. However, since a non-contextual theory of classical physics does not need to make reference to measurements, it is reasonable to expect that for taking a classical limit, one should use a definition of classicality that is independent of any choice of measurements performed on the system. In order to do so, we note that a special case of a quasiprobability representation is one for which every set of POVM elements $\{\hat{E}_i\}$ is specified by a set of non-negative real-valued functions $E^R_i(z)$ i.e. every measurement has a classical representation. In such a representation, any states for which $R(z)\geq 0$ for all $z$ necessarily permit a classical explanation, and we will refer to quantum states which satisfy this property as \textit{effectively classical states}. Providing a measurement-independent notion of classicality, the advantage of this approach is that one may guarantee the effective classicality of dynamics without needing to additionally study the choice of measurements being performed.

To study the representation of measurements, and in particular POVMs, one may ask what form $E_i^Q(q,p)$ and $E_i^P(q,p)$ must take, given the definitions of $Q(q,p)$ and $P(q,p)$ above, in order for equation \eqref{eq: Rep} to hold. By substitution and using the linearity of the trace, it is straightforward to check that the form of POVMs is reversed compared to the states between the two representations: namely that $E_i^P(q,p)$ is obtained by using $\hat{E}_i$ in place of $\hat{\rho}$ in equation \eqref{eq: Q}, while $E_i^Q(q,p)$ is defined implicitly by equation \eqref{eq: P} but with $\hat{E}_i$ in place of $\hat{\rho}$. Just as the $Q$ distribution is positive for all quantum states, this means that the $P$ representations of measurements, $E_i^P(q,p)$, are always positive functions of phase space. Given the definition of effective classicality given above, this means that states which have positive $P$ distributions should be considered to be effectively classical according to this definition.

\section{Relating states and dynamics in the partial quasiprobability representations} \label{app: W,P,Q}

A well known property of the three common quasiprobability distributions is that they may be related via convolution with a Gaussian (also known as a Weierstrass transform). Specifically, the Wigner distribution $W$ may be obtained from the Glauber-Sudarshan $P$ distribution by a convolution with a Gaussian with variance $\frac{1}{2}\hbar s^2$ in $q$ and $\frac{1}{2}\hbar s^{-2}$ in $p$, and in turn the Husimi $Q$ representation may be obtained from the Wigner representation by the same convolution \cite{gardiner2004quantum,wiseman_milburn_2009,takahashi1986wigner}. These relations are unchanged by when one considers instead the partial quasiprobability representations $\varrho^W, \varrho^P, \varrho^Q$, and so using the differential operator representation \cite{hirschman2012convolution} of the convolution 
\begin{equation}
    \mathcal{D}^\frac{1}{2}=e^{\frac{1}{2}(\frac{\hbar s^2}{2} \frac{\partial^2}{\partial q^2}+\frac{\hbar}{2s^2}\frac{\partial^2}{\partial p^2})},
\end{equation}
we may write them as
\begin{equation}
\begin{split}
\varrho^W(q,p)&=\mathcal{D}^\frac{1}{2}\varrho^P(q,p)\\
\varrho^Q(q,p)&=\mathcal{D}^\frac{1}{2}\varrho^W(q,p).
\end{split}
\end{equation} For the different representations to be all equivalent, the mapping between the quasiprobability distributions must be bijective, and thus the convolutions must be invertible. While this is not possible for general functions on phase space \cite{hirschman2012convolution}, in this case it is possible on the restricted domain formed by the sets of all possible partial Husimi and Wigner distributions \cite{takahashi1989distribution}. In terms of the differential operator $\mathcal{D}$, these inverse maps may be written in terms of the differential operator
\begin{equation}
    \mathcal{D}^{-\frac{1}{2}}=e^{-\frac{1}{2}(\frac{\hbar s^2}{2} \frac{\partial^2}{\partial q^2}+\frac{\hbar}{2s^2}\frac{\partial^2}{\partial p^2})}
\end{equation} which gives
\begin{equation}
\begin{split}\varrho^P(q,p)&=\mathcal{D}^{-\frac{1}{2}}\varrho^W(q,p)\\
\varrho^W(q,p)&=\mathcal{D}^{-\frac{1}{2}}\varrho^Q(q,p).
    \end{split}
\end{equation}
Having specified the maps between states in the three representations, one may construct the dynamics in any representation from another by mapping the state, evolving in that representation, and then mapping back to the original representation. In particular, using the form of the generator in the partial Wigner representation $\mathcal{L}^W$, given in \eqref{eq: L_W}, one may construct generators for $\mathcal{L}^Q$ and $\mathcal{L}^P$, which take the form
\begin{equation}
    \mathcal{L}^Q=\mathcal{D}^\frac{1}{2}\mathcal{L}^W \mathcal{D}^{-\frac{1}{2}} = e^{\ad{\frac{\hbar s^2}{4} \frac{\partial^2}{\partial q^2}+\frac{\hbar}{4s^2} \frac{\partial^2}{\partial p^2}}}\mathcal{L}^W
\end{equation}
\begin{equation}
    \mathcal{L}^P=\mathcal{D}^{-\frac{1}{2}}\mathcal{L}^W \mathcal{D}^{\frac{1}{2}} = e^{-\ad{\frac{\hbar s^2}{4} \frac{\partial^2}{\partial q^2}+\frac{\hbar}{4s^2} \frac{\partial^2}{\partial p^2}}}\mathcal{L}^W,
\end{equation} where we have used the relation $e^{\ad{\mathcal{B}}}\mathcal{A}=e^{\mathcal{B}}\mathcal{A}e^{-\mathcal{B}}$. To compute the generators to $O(1)$ in $\hbar$, one can use the definition of the exponential of the adjoint, and the generator of $\mathcal{L}^W$, and expand in orders of $\hbar$. Taking first the generator of partial Husimi dynamics, we find
\begin{widetext}
\begin{equation}
\begin{split}
    \mathcal{L}^Q&=\bigg[1+\ad{\frac{\hbar s^2}{4} \frac{\partial^2}{\partial q^2}+\frac{\hbar}{4s^2} \frac{\partial^2}{\partial p^2}} + \frac{1}{2}\ad{\frac{\hbar s^2}{4} \frac{\partial^2}{\partial q^2}+\frac{\hbar}{4s^2} \frac{\partial^2}{\partial p^2}}^2 + \ldots \bigg]\ \ \bigg( -\frac{i}{\hbar}[H,\dt] +\frac{1}{2}(\{H,\dt\} - \{\dt,H\}) + \ldots  \bigg) \\
    &=-\frac{i}{\hbar}[H,\dt] +\frac{1}{2}(\{H,\dt\} - \{\dt,H\}) + \ad{\frac{\hbar s^2}{4} \frac{\partial^2}{\partial q^2}+\frac{\hbar}{4s^2} \frac{\partial^2}{\partial p^2}}\bigg(-\frac{i}{\hbar}[H,\dt]\bigg) + O(\hbar).
\end{split}
\end{equation} Computing the adjoint action explicitly gives 
\begin{equation}
    \mathcal{L}^Q\big\rvert_{O(\hbar^0)}=-\frac{i}{\hbar}[H,\dt] +\frac{1}{2}(\{H,\dt\} - \{\dt,H\}) -\frac{is^2}{2}[\frac{\partial H}{\partial q},\frac{\partial \dt}{\partial q}] -\frac{i}{2s^2}[\frac{\partial H}{\partial p},\frac{\partial \dt}{\partial p}] -\frac{is^2}{4}[\frac{\partial^2 H}{\partial q^2},\dt] -\frac{i}{4s^2}[\frac{\partial^2 H}{\partial p^2},\dt], 
\end{equation} as given in \eqref{eq: L_Q}. Similarly, one may compute the same for the partial Glauber-Sudarshan dynamics, which differs only by a minus sign, giving
\begin{equation}
    \mathcal{L}^P\big\rvert_{O(\hbar^0)}=-\frac{i}{\hbar}[H,\dt] +\frac{1}{2}(\{H,\dt\} - \{\dt,H\}) +\frac{is^2}{2}[\frac{\partial H}{\partial q},\frac{\partial \dt}{\partial q}] +\frac{i}{2s^2}[\frac{\partial H}{\partial p},\frac{\partial \dt}{\partial p}] +\frac{is^2}{4}[\frac{\partial^2 H}{\partial q^2},\dt] +\frac{i}{4s^2}[\frac{\partial^2 H}{\partial p^2},\dt],
\end{equation} as in \eqref{eq: L_P}.
\end{widetext}
\bibliography{refsQQCQ}

\end{document}